\documentclass[12pt]{article}
\usepackage{amsmath}
\usepackage{graphicx,psfrag,epsf}
\usepackage{enumerate}
\usepackage{enumitem}
\usepackage{natbib}
\usepackage[pdfstartview=FitV]{hyperref}

\usepackage{amsthm}
\usepackage{amssymb}
\usepackage{bm}
\usepackage{bbm}
\usepackage{dsfont}
\usepackage{caption}
\usepackage[labelformat=simple]{subcaption}

\usepackage{tabularx}
\usepackage{tabu}
\usepackage{pdfpages}
\usepackage{array}
\usepackage{longtable}
\usepackage{pdflscape}

\usepackage{algorithm}
\usepackage{algpseudocode}

\newcolumntype{L}[1]{>{\raggedright\let\newline\\\arraybackslash\hspace{0pt}}m{#1}}
\newcolumntype{C}[1]{>{\centering\let\newline\\\arraybackslash\hspace{0pt}}m{#1}}
\newcolumntype{R}[1]{>{\raggedleft\let\newline\\\arraybackslash\hspace{0pt}}m{#1}}

\usepackage{url}
\DeclareMathOperator*{\argmax}{arg\,max}
\DeclareMathOperator*{\argmin}{arg\,min}

\usepackage{tikz}
\tikzset{
box/.style = {shape=rectangle, 
align=center}
}

\newcommand{\beginsupplement}{%
\setcounter{table}{0}
\renewcommand{\thetable}{S\arabic{table}}%
\setcounter{figure}{0}
\renewcommand{\thefigure}{S\arabic{figure}}%
\setcounter{subsection}{0}
\renewcommand{\thesubsection}{S\arabic{subsection}}%
\setcounter{equation}{0}
\renewcommand{\theequation}{S\arabic{equation}}%
}

\theoremstyle{plain}

\theoremstyle{definition}

\theoremstyle{remark}

\newcommand{\blind}{0}

\newcommand{\T}{\text{T}}
\newcommand{\data}{\text{data}}

\newcommand{\U}{\text{U}}
\newcommand{\E}{\text{E}}
\newcommand{\OO}{\text{O}}
\newcommand{\bp}{\bm{p}}

\newcommand{\bw}{\bm{w}}

\newcommand{\Dir}{\text{Dir}}
\newcommand{\Beta}{\text{Beta}}
\newcommand{\TBeta}{\text{TBeta}}

\newcommand{\Weibull}{\text{Weibull}}

\newcommand{\D}{\mathcal{D}}

\newcommand{\A}{\mathcal{A}}
\newcommand{\M}{\mathcal{M}}

\newcommand{\I}{\mathcal{I}}
\newcommand{\dd}{\text{d}}
\newcommand{\bone}{\mathbbm{1}}

\begin{document}

\def\spacingset#1{\renewcommand{\baselinestretch}%
{#1}\small\normalsize} \spacingset{1}

%%%%%%%%%%%%%%%%%%%%%%%%%%%%%%%%%%%%%%%%%%%%%%%%%%%%%%%%%%%%%%%%%%%%%%%%%%%%%%

\if0\blind
{
  \title{\bf PoD-TPI: Probability-of-Decision Toxicity Probability Interval Design to Accelerate Phase I Trials}
  \author{Tianjian Zhou\thanks{Department of Public Health Sciences, The University of Chicago, Chicago, USA. E-mail: \href{mailto:tjzhou@uchicago.edu}{tjzhou@uchicago.edu}} , Wentian Guo\thanks{Laiya Consulting, Inc., Chicago, USA} , and Yuan Ji\footnotemark[1] }
  \maketitle
} \fi

\if1\blind
{
  \bigskip
  \bigskip
  \bigskip
  \begin{center}
    {\LARGE\bf Title}
\end{center}
  \medskip
} \fi

\bigskip
\begin{abstract}
Cohort-based enrollment can slow down dose-finding trials since the outcomes of the previous cohort must be fully evaluated before the next cohort can be enrolled.
This results in frequent suspension of patient enrollment.
The issue is exacerbated in recent immune oncology trials where toxicity outcomes can take a long time to observe.
We propose a novel phase I design, the probability-of-decision toxicity probability interval (PoD-TPI) design, to accelerate phase I trials. 
PoD-TPI enables dose assignment in real time in the presence of pending toxicity outcomes. 
With uncertain outcomes, the dose assignment decisions are treated as a random variable, and we calculate the posterior distribution of the decisions. 
The posterior distribution reflects the variability in the pending outcomes and allows a direct and intuitive evaluation of the confidence of all possible decisions. 
Optimal decisions are calculated based on 0-1 loss, and extra safety rules are constructed to enforce sufficient protection from exposing patients to risky doses.
A new and useful feature of PoD-TPI is that it allows investigators and regulators to balance the trade-off between enrollment speed and making risky decisions by tuning a pair of intuitive design parameters.
Through numerical studies, we evaluate the operating characteristics
of PoD-TPI and demonstrate that PoD-TPI shortens trial duration and
maintains trial safety and efficiency compared to existing time-to-event designs. 
\end{abstract}

\noindent%

\spacingset{1.45}

\section{Introduction}
\label{sec:intro}

Phase I clinical trials in oncology aim to find the maximum tolerated dose (MTD) of a new drug. The MTD is defined as the highest dose with a dose limiting toxicity (DLT) probability close to or lower than a pre-specified target level $p_{\T}$. Many phase I dose-finding designs like the 3+3 design \citep{storer1989design} and the continual reassessment method (CRM, \citealp{o1990continual}) have been proposed for this purpose, along with the goal of maximizing the chance of treating patients at safe and efficacious doses.
In practice, usually a set of candidate dose levels (from low to high) $\{ 1, \ldots, D \}$ is pre-specified, and both the toxicity and the efficacy of the drug are assumed to monotonically increase with dose level. 
Let $p_d$ denote the true and unknown DLT probability of dose $d$,  $p_1 \leq \cdots \leq p_D$.
The value of $p_d$ is estimated according to the DLT outcomes of the treated patients, and the selection of the MTD is based on the estimate of $p_d$.

In practical trials, due to limited sample size, the DLT probability $p_d$ is estimated with uncertainty. To account for the uncertainty, a class of interval-based dose-finding designs has been proposed.
\cite{cheung2002simple} introduced the idea of indifference interval, where any dose with $p_d \in [p_{\T} - \epsilon, p_{\T} + \epsilon]$ can be treated as an estimated MTD for a small value $\epsilon$.
Later, \cite{ji2007dose} proposed a phase I design that solely depends on properties of toxicity probability intervals (TPI), known as the TPI design.
The TPI design was further extended and polished by \cite{ji2010modified}, \cite{ji2013modified} and \cite{guo2017bayesian}, leading to the modified TPI (mTPI) and mTPI-2 designs.
%The basic idea of the TPI designs is to calculate the posterior probability of $p_d$ belonging to a particular interval.
Simpler model-assisted designs like the cumulative cohort design (CCD, \citealp{ivanova2007cumulative}) and the Bayesian optimal interval design (BOIN, \citealp{liu2015bayesian}) also uses intervals as the basis for decision making. The latest i3+3 design \citep{liu2019design} is again based on interval rules, but without a model.

The last decade has seen a growth of the use of immunotherapies for treating cancer.
The toxicities of immunotherapies can be late-onset \citep{kanjanapan2019delayed},
meaning it can take longer time to observe than traditional cytotoxic  cancer therapies.
This presents some challenges to traditional phase I designs like the mTPI-2, as these designs can only make dose decisions when all the previous DLT outcomes are ascertained.
In other words, if some patients are still being followed within their assessment windows,
patient accrual has to be suspended while waiting for the previous outcomes, resulting in waste of precious patient resources and prolonged trial duration.
As a result, there is a need of novel designs that can shorten the
duration of phase I trials. 
Several phase I designs have been proposed to allow dose decisions in
the presence of incomplete DLT outcomes. These include the
time-to-event continual reassessment method (TITE-CRM,
\citealp{cheung2000sequential}), rolling six design
\citep{skolnik2008shortening}, time-to-event Bayesian optimal interval
design (TITE-BOIN, \citealp{yuan2018time}), time-to-event keyboard
design (TITE-keyboard, \citealp{lin2018time}) and rolling TPI design
(R-TPI, \citealp{guo2019rtpi}). Safety of these designs have not
been fully explored and is often a concern in practice. For example,
what is the risk of making a wrong escalation decision when not all
the toxicity outcomes have been observed from existing patients in the
trial? Questions like this need to be carefully addressed to help
investigators and regulators understand the impact of reduction in
trial duration.

In this paper, we propose the probability-of-decision toxicity probability interval (PoD-TPI) design to accelerate phase I trials. 
We calculate the posterior distribution of the number of DLTs among the pending outcomes, with which we then infer the posterior probability of possible dose-finding decisions. 
Importantly, the posterior distribution of decisions enables a direct
and intuitive evaluation of the confidence of the possible decisions, which can be used to assess the risk of any decision rules based on
the posterior distribution.
The decision with the highest PoD is the optimal decision to make.
When the PoD of the optimal decision is high enough, or the optimal decision is not risky (e.g., de-escalation), the decision may be executed without waiting for the pending outcomes. 
On the other hand, if the PoD of the optimal decision is not high enough, and the optimal decision is risky (e.g., escalation), the sensible decision is to suspend the enrollment and wait for some more outcomes to be observed before making a decision.
Through these reasoning,  the proposed PoD-TPI design
quantifies the risk of making aggressive decisions and present them
via simulation studies. Adopting to an optimal and conservative
decision framework and balancing the need to shorten trial duration
and protect patient safety, the PoD-TPI design might be a new and
useful method for practical trials.

The idea of PoD-TPI is different from existing TITE designs.
In TITE-CRM, TITE-BOIN and TITE-keyboard, the follow-up times of the pending patients are incorporated in the likelihood of the DLT probabilities. 
Interest lies in the estimation of $p_d$, and the dose decision is deterministic based on $p_d$.
%Importantly, a direct evaluation of the confidence and risk of the decision is lacking in other TITE designs, making overly aggressive decisions more likely.
In contrast, in PoD-TPI, the follow-up times are used to estimate the probability of a pending patient experiencing DLT in the future. Interest now lies in the posterior distribution of the random number of DLTs if all patients were fully followed, and the dose decision is treated as a random variable.
The idea of PoD-TPI is also different from the R-TPI design. The R-TPI consists of fixed decision rules that accommodate pending outcomes. To ensure safety, dose escalation is only allowed when the mTPI-2 decision is still escalation even if all pending outcomes are DLTs.
However, R-TPI does not make use of the follow-up times thus is less efficient.
In addition to the modeling and application of PoDs, PoD-TPI has two other unique features: (1) the use of probability cutoffs to calibrate the risks to patients, and (2) the proposed risk as a metric for assessing the operating characteristics of designs that do not use complete data. These features will be more clear in the following discussion.

The rest of the paper is organized as follows. In Section \ref{sec:mtpi2}, we review the mTPI-2 design, upon which the PoD-TPI design is built. In Section \ref{sec:titetpi}, we propose the PoD-TPI design based on a probability model for the time to DLT, which is developed in Section \ref{sec:prob_model}. 
In Section \ref{sec:trial_example}, we illustrate the PoD-TPI design through two hypothetical trial examples.
In Section \ref{sec:simulation}, we conduct extensive simulation studies to evaluate the operating characteristics of the PoD-TPI design. In Section \ref{sec:discussion}, we conclude with a discussion.

\section{Review of mTPI-2}
\label{sec:mtpi2}

We first give a brief review of the mTPI-2 design \citep{guo2017bayesian}, upon which the PoD-TPI design is anchored.
Under the mTPI-2 design, patients are usually enrolled in cohorts, and the first cohort is treated at the lowest dose.
At a certain moment in the trial, suppose dose $d$ is currently used to treat patients, and denote by $p_d \in [0, 1]$ its true (and unknown) DLT probability. 
After the toxicity outcomes of the previous cohort are observed,
physicians need to make a decision $\A$ for the treatment of the next cohort.
Here $\A \in \{ -1, 0, 1 \}$, corresponding to a decision of de-escalation ($-1$) to the next lower dose, stay ($0$) at the current dose, or escalation (1) to the next higher dose, respectively.
Accordingly, the dose $(d + \A)$ will be used to treat the next cohort, subject to some additional safety rules.
In the case that $d$ is the lowest dose and $\A = -1$, or $d$ is the highest dose and $\A = 1$, the dose $d$ will be kept for the next cohort, i.e., $\A = 0$.

In mTPI-2, the dose decision $\A$ is based on posterior inference of $p_d$.
Suppose a total of $N$ patients have been treated. We use $Y_i$ and $Z_i$ to denote the DLT outcome and dose assignment of patient $i$, where $Y_i = 1$ (or $0$) indicates a DLT (or non-DLT) outcome, and $Z_i \in \{ 1, \ldots, D \}$, $i = 1, \ldots, N$.
Here, a DLT outcome means the patient experiences DLT within some assessment window. If the patient does not experience DLT during the time window, we declare that the patient has no DLT.
The DLT outcome is modeled by a Bernoulli distribution,
\begin{align*}
Y_i \mid Z_i = d \sim \text{Bernoulli}(p_d).
\end{align*}
Let $n_d$ represent the number of patients who have experienced DLT at dose $d$, and let $m_d$ denote those who have not. That is, $n_d = \sum_{i = 1}^N \bone(Y_i = 1, Z_i = d)$ and $m_d = \sum_{i = 1}^N \bone(Y_i = 0, Z_i = d)$. The likelihood of $p_d$ is
\begin{align}
L(p_d \mid n_d, m_d) = p_d^{n_d} (1 - p_d)^{m_d}.
\label{eq:binomial_likelihood}
\end{align}

Suppose the $(N+1)$-th patient is just enrolled and the DLT outcomes for the previous patients are fully observed. To make a proper dose assignment for the new patient, mTPI-2 considers a partition of the parameter space of $p_d$, the $[0, 1]$ interval, into the following three intervals,
\begin{alignat*}{2}
&I_{\E} = [p_{\T} - \epsilon_1, p_{\T} + \epsilon_2] \quad   &&\text{(equivalence interval)},\\
&I_{\U} = [0, p_{\T} - \epsilon_1) \quad  &&\text{(underdosing interval)}, \\
&I_{\OO} = (p_{\T} + \epsilon_2, 1] \quad  &&\text{(overdosing interval)}.
\end{alignat*}
The values $\epsilon_1$ and $\epsilon_2$ are elicited by physicians, such that the doses within $I_{\E}$ are considered so close to the MTD that the physicians would agree to select them as the estimated MTD.
The lengths of the three intervals $I_{\E}$, $I_{\U}$ and $I_{\OO}$ are different. To avoid undesirable decisions due to the effect of Ockham's razor \citep{jefferys1992ockham}, the intervals $I_{\U}$ and $I_{\OO}$ are further divided into subintervals $\{ I_{\U_1}, \ldots, I_{\U_{K_1}}\}$ and 
$\{ I_{\OO_1}, \ldots, I_{\OO_{K_2}}\}$.
The subintervals have an equal length of $(\epsilon_1 + \epsilon_2)$, except those reaching the boundary of $[0, 1]$.
For example, when $p_{\T} = 0.3$ and $\epsilon_1 = \epsilon_2 = 0.05$, the equivalence interval is $I_{\E} = [0.25, 0.35]$. The underdosing interval is divided into $K_1 = 3$ subintervals,
$I_{\U_1} = [0.15, 0.25)$, $I_{\U_2} = [0.05, 0.15)$ and $I_{\U_3} = [0, 0.05)$, and the overdosing interval is divided into $K_2 = 7$ subintervals, $I_{\OO_1} = (0.35, 0.45]$, $I_{\OO_2} = (0.45, 0.55]$, $I_{\OO_3} = (0.55, 0.65]$, $I_{\OO_4} = (0.65, 0.75]$, $I_{\OO_5} = (0.75, 0.85]$, $I_{\OO_6} = (0.85, 0.95]$ and $I_{\OO_7} = (0.95, 1]$.

Finally, mTPI-2 casts the dose-finding decision as a model selection problem under a decision theoretic framework. Let model $\M_d = j \in \{\U_1$, $\ldots$, $\U_{K_1}$, $\E$, $\OO_1$, $\ldots$, $\OO_{K_2} \}$ be an indicator of $p_d \in I_{j}$.
Each candidate model is assigned a prior probability, and a prior distribution is constructed for $p_d$ given $\M_d$ being the true model. Specifically,
\begin{align*}
&\Pr(\M_d = j) = 1/(K_1 + K_2 + 1), \; \text{for $j = \U_1, \ldots, \U_{K_1}, \E, \OO_1, \ldots, \OO_{K_2}$};  \\
&p_d \mid \mathcal{M}_d \sim \TBeta(1, 1; I_{\mathcal{M}_d}),
\end{align*}
where $\TBeta(\cdot, \cdot; I)$ represents a truncated beta distribution restricted to interval $I$. 
Let $\M^* = \argmax_j \Pr(\M_d = j \mid n_d, m_d)$.
The decision rule of mTPI-2 is given by
\begin{align*}
\A = 
\begin{cases}
0, \quad &\text{if $\M^* = \E$}; \\
1, \quad &\text{if $\M^* \in \{ \U_1, \ldots, \U_{K_1} \}$}; \\
-1, \quad &\text{if $\M^* \in \{ \OO_1, \ldots, \OO_{K_2} \}$}; \\
\end{cases}
\end{align*}
It is shown in \cite{guo2017bayesian} that the decision rule is optimal in the sense that it minimizes the posterior expected loss under a 0-1 loss.

Suppose $p_{\T}$, $\epsilon_1$ and $\epsilon_2$ are given and fixed. The decision $\A$ only depends on the values of $n_d$ and $m_d$. As a result, we can write $\A = \A(n_d, m_d): \mathbb{N} \times \mathbb{N} \rightarrow \{ -1, 0, 1 \}$ as a deterministic function of $n_d$ and $m_d$. For example, with $p_{\T} = 0.30$ and $\epsilon_1 = \epsilon_2 = 0.05$, we have $\A(0, 3) = 1$, $\A(1, 2) = 0$, $\A(2, 1) = -1$ and $\A(3, 0) = -1$.
The notation of $\A(n_d, m_d)$ will be used in the PoD-TPI design next.

\section{The PoD-TPI Design}
\label{sec:titetpi}

As noted in Section \ref{sec:intro}, a potential limitation of the mTPI-2 design (and many other cohort-based phase I designs) is that the toxicity outcomes for the previous cohorts must be fully observed before a dose decision can be made for the next cohort.
While waiting for the DLT assessment of the previous patients, patient accrual needs to be suspended and eligible patients have to be turned away. Such trial suspension is undesirable in practice.
The PoD-TPI design is motivated by the need to reduce the frequency of enrollment suspension but at the same time maintain safety.

Under the PoD-TPI design, patients are enrolled in cohorts, say, of size 3.
The first cohort of patients is treated at the starting dose level, denoted by $d_0$. In most cases, $d_0$ is chosen as the lowest dose level, $d_0 = 1$. 
%To characterize the pending outcomes, let $\tau$ denote the length of the DLT assessment window.
%In oncology, $\tau$ is usually 21 or 28 days, corresponding to a cycle of treatment.

At a given moment of the trial, suppose a total of $N$ patients have been treated, the current dose is $d$, and the $(N+1)$-th patient is available for enrollment.
Recall that $Y_i$ and $Z_i$ denote the DLT outcome and dose assignment of patient $i$, respectively, $i = 1, \ldots, N$.
In particular, $Y_i = 1$ (or 0) represents that patient $i$ experiences (or does not experience) DLT within the assessment window.
Since patients enter clinical trials at random time, it is often the case that when the $(N + 1)$-th patient is eligible for enrollment, some previously enrolled patients are still being followed without definitive DLT outcomes, thus their DLT outcomes $Y_i$'s are unknown.
Let $B_i$ be the indicator for an unknown DLT outcome, where $B_i = 1$ (or $0$) denotes that the DLT outcome of patient $i$ is  unknown (or observed).
We write $n_d = \sum_{i = 1}^N \bone(Y_i = 1, Z_i = d, B_i = 0)$ and $m_d = \sum_{i = 1}^N \bone(Y_i = 0, Z_i = d, B_i = 0)$ the numbers of patients with observed DLTs and non-DLTs, respectively.
In addition, we use $r_d = \sum_{i = 1}^N \bone(Z_i = d, B_i  = 1)$ to denote the number of patients with pending outcomes and write $\I_d = \{ i: Z_i = d, B_i  = 1 \}$ the index set of these patients.
Lastly, we denote $S_d = \sum_{i = 1}^N \bone(Y_i = 1, Z_i = d, B_i = 1)$ the number of DLTs among the $r_d$ pending patients that would have been seen had these patients finished their DLT assessment.
Since these patients are still being followed, $\{ Y_i: i \in
\I_d \}$ are not observed and are random variables, and so as
$S_d$. We have $S_d \in \{ 0, 1, \ldots, r_d \}$. 
The following figure summarizes the patient statistics at dose $d$.

\begin{center}
\begin{tikzpicture}
[
  grow                    = right,
  sibling distance        = 1.8em,
  level distance          = 10em,
  edge from parent/.style = {draw, -latex},
  % every node/.style       = {font=\footnotesize},
  sloped
]
\node [box] {Total  $(n_d + m_d + r_d)$}
    child { node [box] {$r_d$ pending} 
        child { node [box] {$(r_d - S_d)$ non-DLTs} edge from parent}
        child { node [box] {$S_d$ DLTs (random)} edge from parent}
    edge from parent}
    child { node [box] {$m_d$ non-DLTs} edge from parent}
    child { node [box] {$n_d$ DLTs} edge from parent};
\end{tikzpicture}
\end{center}

%Through proper statistical modeling, we can calculate $\Pr(S_d = s \mid \data)$.
If $S_d$ is known, then the dose-finding decision is given by $A_d = \A(n_d + S_d, m_d + r_d - S_d)$, where $\A$ is the decision function of mTPI-2 (Section \ref{sec:mtpi2}).
However, since $S_d$ is not observed and random, the decision $A_d$ becomes a random variable too.
Suppose through statistical inference one could calculate the posterior probability $\Pr(S_d = s \mid \data)$, then we define
\begin{align}
\Pr(A_d = a \mid \data) = \sum_{s:  \A(n_d + s, m_d + r_d - s) = a} \Pr(S_d = s \mid \data),
\label{eq:post_decisions}
\end{align}
which gives the PoD for each decision $a \in \{ -1, 0, 1\}$ based on mTPI-2. 
Details about the probability model and statistical inference are described in the next section.
If $d$ is the lowest dose, de-escalation is not possible. Similarly, if $d$ is the highest dose, escalation is not possible.
To ensure that the PoD is well defined, in Supplementary Section \ref{supp:sec:pod} we show that $\Pr(A_d = a \mid \data)$ sums up to 1 over the space of $a$. 
If there is no pending patient at the current dose ($r_d = 0$),  we have $S_d \equiv 0$, and $A_d \equiv \A(n_d, m_d)$, i.e., the mTPI-2 decision.

Next, define $\ell(A_d = a, A_d^{\text{true}} = b)$ the loss of making dose decision $a$ if $b$ is the true dose decision that should have been made. We consider the 0-1 loss,
\begin{align}
\ell(a, b) = 
\begin{cases}
1, \quad \text{if $a \neq b$}; \\
0, \quad \text{if $a = b$},
\end{cases}
\text{for $a, b \in \{ -1, 0, 1\}$.}
\label{eq:loss_function}
\end{align}
Let $A_d^* = \min \{ \argmax_a \Pr(A_d = a \mid \data) \}$ denote the most conservative decision within the set of decisions that have the highest PoD. 
In other words, if $\argmax_a \Pr(A_d = a \mid \data)$ returns a
single decision among $\{ -1, 0, 1\}$, then $A_d^*$ equals that decision. Otherwise, if multiple decisions tie for the highest
PoD, we choose the more conservative one using the ``$\min$''
function. 
It is easy to see that the dose decision $A_d^*$ minimizes the posterior expected loss under the loss function \eqref{eq:loss_function} thus is the optimal decision to make, i.e., Bayes' rule. 

Denote by $\gamma_{d, a} = \Pr(A_d = a \mid \data)$ for $a \in \{
-1, 0, 1 \}$. To ensure the safety of the design, we introduce
two essential suspension rules.

\vspace{0.5em}

\noindent
\fbox{
\begin{minipage}{0.96\textwidth}
  \begin{description}
    \setlength\itemsep{0.1in}
\item[If $A_d^* = 1$, i.e. escalation:] We suspend the trial if
(i)  $\gamma_{d, 1} < \pi_{\E}$ for some threshold $\pi_{\E} \in [0.33, 1]$ or 
(ii) $m_d = 0$.
Condition (i) reflects that escalation is not allowed if the confidence of escalation is less than $\pi_{\E}$.
A larger $\pi_{\E}$ represents more conservative dose escalations.
Condition (ii)  means escalation is not allowed until at least one patient at the current dose has finished the DLT assessment and does not experience DLT, similar to the rule in  \cite{normolle2006designing}.  
\item[If $A_d^* = 0$, i.e. stay:] We suspend the trial if $\gamma_{d, -1} > \pi_{\text{D}}$ for some threshold $\pi_{\text{D}} \in [0, 0.5]$. This means stay is not allowed if there is a relatively high chance of de-escalation.
A smaller $\pi_{\text{D}}$ represents more conservative stays.
\end{description}
\end{minipage}
}
\vspace{0.5em}

If none of the suspension rule is triggered, the optimal decision $A_d^*$ is made.
In real applications, the values $\pi_{\E}$ and $\pi_{\text{D}}$ should be chosen according to the desired extent of safety. 
To ensure safety, we recommend choosing $\pi_{\E} \geq 0.8$ and setting $\pi_{\text{D}} \leq 0.25$.
For example, in the simulation studies (Section \ref{sec:simulation}), we use $\pi_{\E} = 1$ and $\pi_{\text{D}} = 0.15$, which eliminate the chance of risky escalations.
We will discuss more about the important role of $\pi_{\E}$ and $\pi_{\text{D}}$ in Section \ref{sec:simulation}. 
The dose assignment rules of PoD-TPI is summarized in Algorithm \ref{alg:dose-assignment}.

\begin{algorithm}[h]
\caption{Dose assignment rule of PoD-TPI. Current dose level is $d$.}
\label{alg:dose-assignment}
\begin{algorithmic}[1]
\If{$r_d = 0$}
\State Assign the patient to dose $d + \A(n_d, m_d)$
\ElsIf{$r_d > 0$}
\State Calculate $\gamma_{d, a} = \Pr(A_d = a \mid \data)$ and $A_d^*
= \min\{\argmax_a \Pr(A_d = a \mid \data)\}$
\If{$A_d^* = 1$}
\If{$\gamma_{d, 1} < \pi_{\E}$ \textbf{or} $m_d = 0$}
\State Suspend accrual
\Else
\State Assign the patient to $(d + 1)$
\EndIf
\ElsIf{$A_d^* = 0$}
\If{$\gamma_{d, -1} > \pi_{\text{D}}$}
\State Suspend accrual
\Else
\State Assign the patient to $d$
\EndIf
\ElsIf{$A_d^* = -1$}
\State Assign the patient to $(d - 1)$
\EndIf
\EndIf
\end{algorithmic}
\end{algorithm}

For practical concerns, similar to existing designs (for example, \citealp{ji2010modified} and \citealp{yuan2018time}), we include the following two safety rules in PoD-TPI throughout the trial.
\begin{description}
\item[\underline{Safety rule 1 (early termination)}:] At any moment in the trial, if $n_1 + m_1 \geq 3$ and $\Pr(p_1 > p_{\T} \mid n_1, m_1) > 0.95$, terminate the trial due to excessive toxicity;
\item[\underline{Safety rule 2 (dose exclusion)}:] At any moment in the trial, if  $n_d + m_d \geq 3$ and $\Pr(p_d > p_{\T} \mid n_d, m_d) > 0.95$, suspend dose $d$ and higher doses from the trial. 
\end{description}
In other words, if a sufficient number ($\geq 3$) of patients has been
treated at dose $d$ with observed outcomes, and these outcomes suggest
$d$ is deemed higher than the MTD, then this dose and higher doses are excluded from the trial.
The posterior probability $\Pr(p_d > p_{\T} \mid n_d, m_d)$ is calculated using the observed data at dose $d$ with a prior distribution $p_d \sim \Beta(1, 1)$.
It is possible that a dose $d$ and higher doses are excluded when some outcomes at $d$ are pending. In this case, we allow these doses to be re-included if the pending outcomes are observed later and the safety rule is no longer violated.
If safety rule 1 is triggered while some outcomes at the lowest dose are pending, we suspend the trial. The trial is resumed if the pending outcomes are observed later and the safety rule is no longer violated, and the trial is permanently terminated otherwise.

%\subsection{Selection of the MTD}
The trial is completed if the number of enrolled patients reaches the pre-specified maximum allowable sample size $N^*$ or  safety rule 1 is triggered.
The last step is to recommend an MTD based on the DLT outcomes.
If the trial is terminated due to safety rule 1, no MTD is selected. Otherwise, after DLT assessment for all patients is finished, we use the same procedure as in \cite{ji2007dose} to report an MTD. 
See Supplementary Section \ref{supp:sec:sel_mtd} for details.

\section{Probability Model}
\label{sec:prob_model}

\subsection{Likelihood Construction}

We construct the likelihood function for the observed data, with which we make inference about the distribution of the time to DLT and calculate the posterior distribution of $S_d$ and $A_d$ (Equation \ref{eq:post_decisions}).

We first introduce some additional notation.
Let $\tau$ denote the length of the DLT assessment window. In oncology, $\tau$ is usually 21 or 28 days, corresponding to a cycle of treatment.
Denote by $T_i$ the time to DLT for patient $i$, $i = 1, \ldots, N$; recall that we assume $N$ patients have been treated.
By definition, $Y_i = \bone( T_i \leq \tau)$, because $Y_i$ represents whether patient $i$ experiences DLT within the assessment window.
Conditional on the dose assignments ($Z_i$'s), the $T_i$'s are assumed to be independent and identically distributed with probability density function $f_{T \mid Z}$ and survival function $S_{T \mid Z}$.

Next, the following notations are defined with respect to the time when the $(N + 1)$-th patient is available for enrollment. To simplify notation, we do not explicitly write out the dependency on time.
Let $U_i = \min\{ \tau, e_{N+1} - e_i \}$ denote the potential censoring time for patient $i$, where $e_i$ is the enrollment time for patient $i$, and $(e_{N+1} - e_i)$ is the time between the enrollment time of patient $i$ and the time when the new patient $(N+1)$ becomes available.
Let $V_i = \min \{ T_i, U_i \}$ denote the follow-up time, and let $\delta_i = \bone ( T_i \leq U_i )$ indicate whether the DLT is observed ($\delta_i = 1$) or censored ($\delta_i = 0$).
We note that the case $\{ \delta_i = 1 \}$ corresponds to $\{ Y_i = 1, B_i = 0 \}$, and $\{ \delta_i = 0 \}$ includes $\{ Y_i = 0, B_i = 0 \}$ and $\{ B_i = 1 \}$.

Based on survival modeling (see, e.g., \citealp{klein2006survival}), patients with observed DLTs ($\delta_i = 1$) contribute $f_{T \mid Z}$ to the likelihood, and patients with censored observations ($\delta_i = 0$) contribute $S_{T \mid Z}$ to the likelihood. Therefore, the likelihood function is
\begin{align}
L = \prod_{i = 1}^N \left[ f_{T \mid Z}(v_i \mid z_i)^{\bone(\delta_i = 1)} S_{T \mid Z}(v_i \mid z_i)^{\bone(\delta_i = 0)} \right].
\label{eq:likelihood0}
\end{align}
We define a model for $f_{T \mid Z}(v_i \mid z_i)$ next.

\subsection{Sampling Model for Time to Toxicity}

We assume a parametric distribution for $T_i$ as follows. First, as in the mTPI-2 design, we assume 
\begin{align}
\Pr(T_i \leq \tau \mid Z_i = d, p_d) = \Pr(Y_i = 1 \mid Z_i = d, p_d) = p_d.
\label{eq:pdf_Y}
\end{align}
That is, with probability $p_d$, the DLT for a patient treated by dose $d$ occurs within $(0, \tau]$.

Conditional on $[T_i \leq \tau]$ (i.e., $[Y_i = 1]$), we assume a piecewise uniform distribution for $[T_i \mid Y_i = 1, Z_i = d]$ on the interval $(0, \tau]$. 
That is, we partition $(0, \tau]$ into $K$ sub-intervals $\{ (h_{k-1}, h_k], k = 1, \ldots, K\}$, where $0 = h_0 < h_1 < \cdots < h_K = \tau$. 
For simplicity, we consider $K = 3$ sub-intervals with equal length, $h_k = k \tau / K$ for $k = 0, 1, 2, 3$.
The $k$-th sub-interval is assigned a weight $w_k$, and $\sum_{k = 1}^K w_k = 1$. 
Conditional on $[Y_i = 1, Z_i = d]$, $T_i$ falls into $(h_{k-1}, h_k]$ with probability $w_k$ and follows a uniform distribution within this interval.
The conditional probability density function of $[T_i \mid Y_i = 1, Z_i = d]$ is thus
\begin{align}
f_{T \mid Y, Z} (t \mid Y_i = 1, Z_i = d, \bw) =  w_k \cdot \frac{1}{h_k - h_{k-1}}, \quad \text{for} \; h_{k-1} < t \leq h_k.
\label{eq:pdf_T}
\end{align}
Implicitly in \eqref{eq:pdf_T}, $T_i$ and $Z_i$ are conditionally independent given $[T_i \leq \tau]$, meaning the conditional distribution of the time to DLT is the same across doses.
In other words, the parameter $\bw$ is shared across doses.
As toxicity data are typically sparse in phase I trials, the conditional independence assumption allows borrow of information across doses and helps the estimation of $\bw$.

Next, according to the law of total probability,
\begin{multline*}
f_{T \mid Z}(t \mid Z_i = d, p_d, \bw)  \\
= \sum_{y \in \{ 0, 1 \} } f_{T \mid Y, Z} (t \mid Y_i = y, Z_i = d, \bw) \Pr(Y_i = y \mid Z_i = d, p_d) \\
= p_d \cdot w_k \cdot \frac{1}{h_k - h_{k-1}}, \quad \text{for} \; h_{k-1} < t \leq h_k.
\end{multline*}
Here, $f_{T \mid Y, Z} (t \mid Y_i = 0, Z_i = d, \bw) = 0$ for $t \leq \tau$, since $\{ Y_i = 0 \}$ indicates $\{ T_i > \tau \}$.
The survival function of $T_i$ is
\begin{align*}
S_{T \mid Z} (t \mid Z_i = d, p_d, \bw)  
&=  1 - \int_{0}^t f_{T \mid Z}(v \mid Z_i = d, p_d, \bw) \, \dd v \\ 
&= 1 - p_d \sum_{k = 1}^K w_k \beta(t, k), \quad \text{for} \; t \leq \tau,
\end{align*}
where 
\begin{align*}
\beta(t, k) = 
\begin{cases} 
1, & \text{if }  v > h_k; \\
\frac{t - h_{k-1}}{h_k - h_{k-1}},  & \text{if } v \in (h_{k-1}, h_k], k = 1, \ldots, K; \\
0,  & \text{otherwise.}  
\end{cases}
\end{align*}
%Notice that we are not interested in $f_{T \mid Z}$ and $S_{T \mid Z}$ for $t > \tau$, as the follow-up time is always less than or equal to $\tau$.

Finally, writing out the parametric forms of $f_{T \mid Z}$ and $S_{T \mid Z}$ in Equation \eqref{eq:likelihood0}, we obtain the likelihood of $\bp$ and $\bw$,
\begin{align}
L \triangleq L(\bp, \bw \mid \data)
\propto \prod_{k = 1}^K w_k^{n_{\cdot k}} \prod_{d = 1}^D \left\{ p_d^{n_d}  (1 - p_d)^{m_d} \prod_{i \in \mathcal{I}_d} \left[ 1 - p_d \sum_{k=1}^K w_k \beta(v_i, k) \right] \right\}.
\label{eq:likelihood}
\end{align}
Here, $n_{\cdot k} = \sum_{i : y_i = 1} \bone(h_{k-1} < v_i \leq h_{k})$ is the number of patients (across all doses) who have experienced DLT in the time interval $(h_{k-1}, h_k]$.
The pending patients' follow-up times are included in the likelihood \eqref{eq:likelihood} via $\beta(v_i, k)$. If the DLT outcomes are fully observed, i.e., $\mathcal{I}_d = \varnothing$, the likelihood of $p_d$ reduces to the Bernoulli likelihood  \eqref{eq:binomial_likelihood}.

\subsection{Prior and Posterior}

We complete the probability model with prior models for the parameters $\bp = (p_1, \ldots, p_D)$ and $\bw = (w_1, \ldots, w_K)$. We assume
\begin{align}
p_d \sim \Beta(\theta_{d1}, \theta_{d2}), \;\; \text{and} \;\;
\bw \sim \Dir(\eta_1, \ldots, \eta_K).
\label{eq:prior}
\end{align}
Here, $\theta_{d1}$ and $\theta_{d2}$ can be chosen based on prior guess of the DLT probability of each dose, and $(\eta_1, \ldots, \eta_K)$ can be chosen based on prior knowledge of the time to DLT falling into each sub-interval.
If such information is not available, we recommend simply setting $\theta_{d1} = \theta_{d2} = 1$ and $\eta_1 = \cdots = \eta_K = 1$.

\paragraph{Posterior of $A_d$.}
With the likelihood \eqref{eq:likelihood} and the prior model \eqref{eq:prior}, we can conduct posterior inference on $\bp$ and $\bw$. Specifically,
\begin{align*}
\pi(\bp, \bw \mid \data) \propto \prod_{d = 1}^D \pi_0(p_d) \times \pi_0(\bw) \times L(\bp, \bw \mid \data),
\end{align*}
where $\pi_0(p_d)$ and $\pi_0(\bw)$ are the prior models as in \eqref{eq:prior}.
Markov chain Monte Carlo simulation is used to draw samples from the posterior distribution $\pi(\bp, \bw \mid \data)$.

Based on the sampling models \eqref{eq:pdf_Y} and \eqref{eq:pdf_T}, we can calculate the probability that a patient experiences DLT within the assessment window given the patient has been followed for $v_i$ ($< \tau$) without DLT, i.e., the conditional probability of $\{ Y_i = 1 \}$ for $i \in \I_d$.
Recall that $\I_d$ contains the indices of the pending patients.
For a patient $i \in \I_d$, we have
\begin{align*}
q_i(v_i, d, p_d, \bw) {}&\triangleq \Pr(Y_i = 1 \mid T_i > v_i, Z_i = d, p_d, \bw) \\
{}&= \frac{\Pr(T_i > v_i \mid Y_i = 1, Z_i = d, \bw) \Pr(Y_i = 1 \mid Z_i = d, p_d)}{\sum_{y =0 }^1 \Pr(T_i > v_i \mid Y_i = y, Z_i = d, \bw) \Pr(Y_i = y \mid Z_i = d, p_d)} \\
{}&= \frac{\left[ 1 - \sum_{k = 1}^K w_k \beta(v_i, k) \right] p_d}{\left[1 - \sum_{k = 1}^K w_k \beta(v_i, k) \right] p_d + (1 - p_d)}, \quad (v_i < \tau ).
\end{align*}

Recall that $S_d$ is the number of patients that will experience DLTs among the pending patients at dose $d$. Therefore, mathematically $S_d = \sum_{i \in \I_d} Y_i$. By definition, given the observed data (including the pending patients' follow-up times), $[S_d \mid p_d, \bw, \data]$ follows a Poisson binomial distribution,
\begin{align*}
S_d \mid p_d, \bw, \data \sim \text{Poisson-binomial}(q_i, i \in \I_d).
\end{align*}
Here, the Poisson binomial distribution is the distribution of the sum of independent Bernoulli random variables that not necessarily have the same success probabilities. See, for example, \cite{chen1997statistical} for an introduction.
Furthermore, we have 
\begin{align*}
\Pr(S_d = s \mid \data) =
\int_{\bw} \int_{p_d} \Pr(S_d = s \mid p_d, \bw, \data) \pi(p_d, \bw \mid \data) \dd p_d \, \dd \bw.
\end{align*}
This integral can be approximated using posterior samples of $p_d$ and $\bw$.
Finally, we can calculate $\Pr(A_d = a \mid \data)$ according to Equation \eqref{eq:post_decisions}.

\section{Hypothetical Trial Examples}
\label{sec:trial_example}

\begin{figure}[h!]
\begin{center}
\includegraphics[width=.85\textwidth]{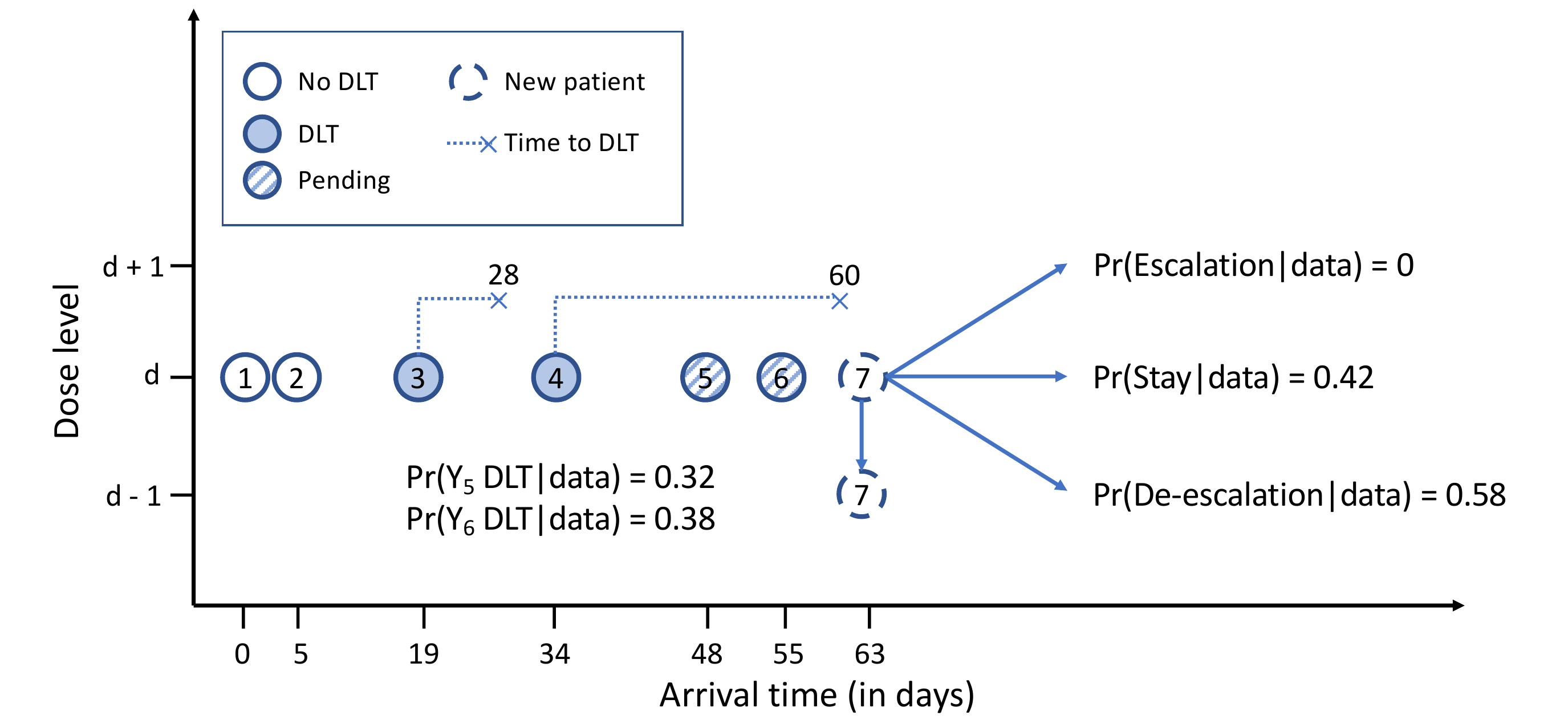}

(a) Hypothetical Trial 1

\includegraphics[width=.85\textwidth]{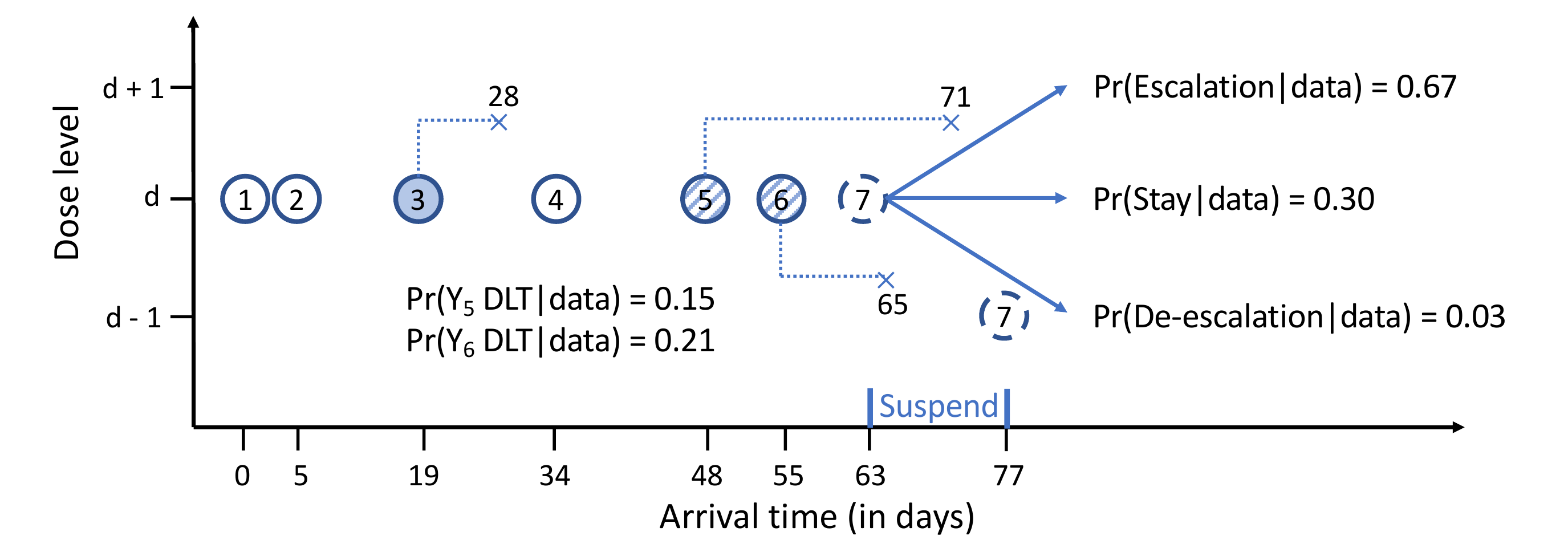}

(b) Hypothetical Trial 2
\end{center}
\caption{Two hypothetical dose-finding trials using the PoD-TPI design. In both trials, by the time patient 7 arrives, patients 1, 2, 3 and 4 have observed outcomes, and patients 5 and 6 are still being followed without definitive outcomes.
In trial 1, based on the posterior probabilities of decisions, patient 7 is assigned to the next lower dose. In trial 2, due to safety concerns, the trial is suspended, and patient 7 is turned away. After the pending outcomes are observed, the next patient is  assigned to the next lower dose. }
\label{fig:illustration}
\end{figure}

We illustrate the PoD-TPI design through two hypothetical trials in Figure \ref{fig:illustration}.
Suppose the target DLT probability is $p_{\T} = 0.3$, and the DLT assessment window is $\tau = 28$ days.
We use $\epsilon_1 = \epsilon_2 = 0.05$ as the bounds of the equivalence interval.
In addition, we set the thresholds $\pi_{\E} = 1$ and $\pi_{\text{D}} = 0.15$ for trial suspension, and let $\theta_{d1} = \theta_{d2} = 1$ and $\eta_1 = \cdots =  \eta_K = 1$.
In the first trial, by the time the new patient 7 arrives on day 63, 6 patients have been treated at dose $d$, among whom patients 1, 2, 3 and 4 have observed outcomes with two DLTs and two non-DLTs ($n_d = 2$ and $m_d = 2$).
The times to DLTs for patients 3 and 4 are $t_3 = 9$ days and $t_4 = 26$ days, respectively.
Patients 5 and 6 are still pending ($r_d = 2$) with follow-up times $v_5 = 15$ days and $v_6 = 8$ days.
Assume $d$ is neither the lowest dose nor the highest dose, thus both de-escalation and escalation are possible.
Via the probability model introduced in Sections \ref{sec:titetpi} and \ref{sec:prob_model}, using
the observed DLT data on patients 1--4 and time-to-event data on patients 5 and 6, PoD-TPI estimates
$\Pr(S_d = 2 \mid \data) = 0.12$, $\Pr(S_d = 1 \mid \data) = 0.46$ and $\Pr(S_d = 0 \mid \data) = 0.42$ and calculates the PoDs $\Pr(A_d = a \mid \data)$, as seen in Figure \ref{fig:illustration}. 
De-escalation ($A_d = -1$) has the highest PoD, and PoD-TPI recommends de-escalating the dose for patient 7 although two patients are still pending. 
In this way, PoD-TPI saves time and avoids trial suspension compared to mTPI-2.
In the second trial, the only difference is that patient 4 does not have a DLT ($n_d = 1$ and $m_d = 3$). In this case, 
$\Pr(S_d = 2 \mid \data) = 0.03$, $\Pr(S_d = 1 \mid \data) = 0.30$ and $\Pr(S_d = 0 \mid \data) = 0.67$. Escalation has the highest PoD, which is 0.67. However, the PoD of escalation does not reach the safety threshold $\pi_{\E}$, meaning there is a chance (0.33) that escalation is not preferred. 
In practice, overly aggressive escalations can cause tremendous risk to patients, thus we use a conservative threshold $\pi_{\E} = 1$ to prevent risky escalations even though the PoD for escalation is 0.67, the largest.
Enrollment is suspended, and patient 7 is turned away. 
Later, a new patient (still labeled as the 7th patient in the trial) arrives on day 77, and by this time it turns out that patients 5 and 6 had DLTs. Therefore, the dose is de-escalated for patient 7. In this way, PoD-TPI avoids the risky escalation and maintains the safety of the trial.

\paragraph{Differences with other TITE designs.}

 The two trial examples highlight the key feature of PoD-TPI: the use of posterior distribution of decisions. 
In the other TITE designs, such as TITE-CRM, TITE-BOIN and TITE-keyboard, the pending outcomes and the follow-up times are incorporated into the survival likelihood \eqref{eq:likelihood}  to help estimate $p_d$.
Dose decision is then based on inference about $p_d$, instead of the PoDs of the decisions. 
For example, TITE-CRM is an extension of the CRM \citep{o1990continual} and models $p_d$ using a parametric curve, $p_d = \phi(d, \alpha)$. Here $\alpha$ is the parameter that defines the dose-toxicity curve, e.g., a regression parameter in a power curve as $p_d = p_{0d}^{\exp(\alpha)}$, where $p_{0d}$'s are pre-specified.
Then, a weighted likelihood for $\alpha$ is constructed through survival modeling,
%Denote by $v_i$ the follow-up time for patient $i$. The weighted likelihood is given by
\begin{align*}
L(\alpha \mid \data) = 
\prod_{i = 1}^N [ \rho_i \phi(z_i, \alpha)]^{\bone(\delta_i = 1)} \left[ 1 - \rho_i \phi(z_i, \alpha) \right]^{\bone(\delta_i = 0)}.
\end{align*}
%\begin{align*}
%L(\alpha \mid n_d, m_d, r_d, \bv) = p_d^{n_d} (1 - p_d)^{m_d} \prod_{i \in \I_d} (1 - \rho_i p_d),
%\end{align*}
%Here $\by = (y_1, \ldots, y_N)$, $\bv = (v_1, \ldots, v_N)$ and $\bz = (z_1, \ldots, z_N)$ are the DLT outcomes, follow-up times and dose assignments for the first $N$ patients, respectively. 
The value $\rho_i$ is the weight of patient $i$, $\rho_i = 1$ if the patient has observed outcome, and $\rho_i = \rho(v_i) \in [0, 1]$ otherwise.
By default, $\rho(v_i) = v_i / \tau$.
TITE-BOIN is an extension of the BOIN design \citep{liu2015bayesian}, which estimates $p_d$ by 
\begin{align*}
\hat{p}_d \propto \frac{n_d + \frac{n_d + 0.5 p_{\T}}{m_d + 1 - 0.5 p_{\T}} \left[r_d - \sum_{i \in \I_d} (v_i / \tau) \right] }{n_d + m_d + r_d}.
\end{align*}
TITE-keyboard is an extension of the keyboard design \citep{yan2017keyboard}, which estimates $p_d$ through the likelihood,
\begin{align*}
L(p_d \mid \data) = p_d^{n_d} (1 - p_d)^{m_d} \prod_{i \in \mathcal{I}_d} (1 - \rho_i p_d).
\end{align*}
Again, $\rho_i$ is the weight of patient $i$.
In either cases, one can conduct inference on $p_d$ and make a dose decision based on the same rules in CRM, BOIN or keyboard.
In contrast, PoD-TPI takes a different approach by making inference on decisions. 
The advantage of PoD-TPI over existing TITE designs is that PoD-TPI  formally takes into account the variability of the pending outcomes and uses probability cutoffs to control the chance of risky decisions.
Consider the second trial example in Figure \ref{fig:illustration}. 
When patient 7 arrives on day 63, using the TITE-CRM, TITE-BOIN or TITE-keyboard designs, based on the estimate of $p_d$, patient 7 will be treated by the next higher dose ($d+1$).
%TITE-BOIN, the estimated $\hat{p}_d = 0.225$, thus the dose will be escalated for patient 7.
However, as the outcomes of patients 5 and 6 turn out to be DLTs later, the escalation decision puts patient 7 in an overly toxic dose.
Such decisions are of major concern to safety committees as they tend to expose patients to unnecessary risks. Some designs take additional steps to suspend enrollment to enhance safety.
For example, TITE-BOIN has a suspension rule when more than 50\% of the outcomes at the current dose are pending, and TITE-keyboard has a suspension rule when fewer than two patients at the current dose level have been completely followed. Nevertheless, as in this example, such suspension rules cannot fully account for the variability of the pending outcomes and the risks of different decisions and would still allow the risky escalation.
%This example illustrates the risk of not taking into account the variability of the pending outcomes.

\section{Numerical Studies}
\label{sec:simulation}

We conduct extensive computer simulations to assess the operating characteristics of the proposed PoD-TPI design.
The implementation of PoD-TPI requires the specification of a few parameters, $\pi_{\E}$, $\pi_{\text{D}}$, $(\theta_{d1}, \theta_{d2})$ and $(\eta_1, \ldots, \eta_K)$, in the suspension rule and prior model.
We use $\pi_{\E} = 1$, $\pi_{\text{D}} = 0.15$, $\theta_{d1} = \theta_{d2} = 1$ and $\eta_1 = \cdots =  \eta_K = 1$.
Note that $\pi_{\E} = 1$ means that escalation must have PoD equal to 1 to be selected as the dose decision when there are pending patients.

We consider the 60 dose-toxicity scenarios reported in \cite{guo2019rtpi}, which cover different target DLT probabilities $p_{\T} \in \{0.10, 0.17, 0.30 \}$, numbers of doses $D \in \{ 3, 4, 5, 6 \}$ and dose-toxicity curves. 
See Supplementary Table \ref{tbl:simu_truth}.
To obtain the mTPI-2 decision rule $\A$, it is necessary to define the equivalence interval. We use $\epsilon_1 = \epsilon_2 = 0.03$ when $p_{\T} = 0.10$, and $\epsilon_1 = \epsilon_2 = 0.05$ when $p_{\T} = 0.17$ or $0.30$.
As in typical phase I oncology trials, we set the length of the DLT assessment period at $\tau = 28$ days.
For a scenario with a total of $D$ doses, we use a maximum sample size of $6D$ patients. We generate the arrival time between two consecutive patients from an exponential distribution with rate $\delta$ per day, meaning the average inter-arrival time is $1 / \delta$ days.
The time to DLT for a patient treated by dose $d$ is generated from a Weibull distribution with shape $\zeta_d$ and scale $\lambda_d$,
\begin{align*}
(T_i \mid Z_i = d) \sim \Weibull(\zeta_d, \lambda_d),
\end{align*}
such that $\Pr(T_i \leq \tau \mid Z_i = d) = p_d$.
To uniquely identify $\zeta_d$ and $\lambda_d$, we impose another constraint, $\Pr(T_i \leq (1 - \gamma) \tau \mid Z_i = d) = (1 - \alpha) p_d$ for some $0 < \gamma < 1$ and $0 < \alpha < 1$, which means if a DLT occurs within the assessment window, with probability $\alpha$ it occurs within the last fraction of $\gamma$.
This leads to
\begin{align*}
\left. \zeta_{d} = \log \left[ \frac{\log(1 - p_d)}{\log(1 - p_d + \alpha p_d)} \right] \middle/ \log \left( \frac{1}{1 - \gamma} \right) \right. , 
\;\; \lambda_{d} = \frac{\tau}{\left[-\log(1 - p_d)\right]^{1/\zeta_{d} } }.
\end{align*}
For each dose-toxicity scenario, we consider the following four settings that cover different accrual rates and DLT time profiles.
\begin{description}
\item[Setting 1:] inter-arrival time is 10 days, and $50\%$ of DLTs occur in the second half of the assessment window, meaning  $\delta = 0.1$, $\alpha = 0.5$ and $\gamma = 0.5$;
\item[Setting 2:] inter-arrival time is 5 days, and $50\%$ of DLTs occur in the second half of the assessment window, meaning  $\delta = 0.2$, $\alpha = 0.5$ and $\gamma = 0.5$;
\item[Setting 3:] inter-arrival time is 10 days, and $80\%$ of DLTs occur in the last quarter of the assessment window, meaning  $\delta = 0.1$, $\alpha = 0.8$ and $\gamma = 0.25$.
\item[Setting 4:] inter-arrival time is 5 days, and $80\%$ of DLTs occur in the last quarter of the assessment window, meaning  $\delta = 0.1$, $\alpha = 0.8$ and $\gamma = 0.25$.
\end{description}
We do not directly consider a setting with a different length of assessment window, as it is equivalent to varying the accrual rate after rescaling the time. Finally, for each scenario and accrual rate and DLT time setting, we simulate 1,000 trials using PoD-TPI as the design.

The performance of the PoD-TPI design is first evaluated based on its reliability and safety. Specifically, the reliability can be measured by the percentage of trials that the MTD is correctly selected (PCS), and the safety can be measured by the average percentages of patients that are correctly allocated to the MTD (percentage of correct allocation, PCA) and are allocated to doses higher than the MTD (percentage of overdosing allocation, POA).
Furthermore, we calculate the percentage of trials concluding a dose above the true MTD (percentage of overdosing selection, POS) and the average percentage of toxicity outcomes (POT).
The average trial duration is also reported.

In addition to the metrics mentioned above, as noted in Section \ref{sec:trial_example}, a major concern for designs allowing patient enrollment with pending DLT information is the possibility of making an \emph{inconsistent} decision, which is another important aspect of reliability and safety. 
Here, an inconsistent decision refers to a decision that is different from what would be made if complete outcomes were observed.
There are six types of inconsistent decisions: 
(1) should de-escalate but stayed (DS), (2) should de-escalate but escalated (DE), (3) should stay but escalated (SE), (4) should stay but de-escalated (SD), (5) should escalate but de-escalated (ED), and (6) should escalate but stayed (ES). DE and SE are the most risky decisions, as they can expose patients to overly toxic doses. DS is also risky for the same reason.
SD, ED and ES are conservative decisions, which can allocate too many patients to sub-therapeutic doses  but are less severe than the overly aggressive decisions.

The performance of PoD-TPI is benchmarked against mTPI-2, which can be seen as a performance upper bound. In addition, we implement R-TPI, TITE-CRM and TITE-BOIN for comparison. 
Specifically, for TITE-CRM, we use the empiric model, $p_d = \phi(d, \alpha) = p_{0d}^{\exp(\alpha)}$. The skeleton $p_{0d}$'s are chosen based on \cite{lee2009model} using an initial MTD guess of $\lceil D/2 \rceil$; the prior for $\alpha$ is $\text{N}(0, 1.34^2)$.
The mTPI-2 can only make dose assignments with complete DLT outcomes, thus we suspend accrual after treating each cohort of patients until all DLT outcomes are fully observed.
By default, for R-TPI, the trial is suspended if more than 3 patients at the current dose are still being followed without outcomes. For TITE-CRM, available patients are always enrolled without suspension. For TITE-BOIN, the trial is suspended if more than half of the outcomes at the current dose are pending.
The same safety rules (Section \ref{sec:titetpi}) are applied to all designs. Also, dose escalation is prohibited until at least one patient at the current dose has finished DLT assessment without experiencing toxicity \citep{normolle2006designing}.

The simulation results are summarized in Tables \ref{tbl:simu_result} and \ref{tbl:simu_inconsist_dec}, which show averages over the 60 dose-toxicity scenarios.
The scenario-specific operating characteristics are reported in Supplementary Figures \ref{supp:fig:pcs} and \ref{supp:fig:inconsist}.
The PCS, PCA, POA, POS and POT of PoD-TPI, reported in Table \ref{tbl:simu_result}, are comparable to those of mTPI-2. On average, the trial duration is shortened by 70 days using PoD-TPI compared to mTPI-2.
On the other hand, R-TPI, TITE-CRM and TITE-BOIN also yield performances generally comparable to those of mTPI-2.
We note that the duration of the trial solely depends on the suspension rules, i.e. how many times an available patient is turned away. As a result, TITE-CRM achieves the optimal trial speed as it does not have a suspension rule. However, the reliability of TITE-CRM is decreased when accrual is fast (Settings 2 and 4).

\begin{table}[ht!]
\caption{Performance of PoD-TPI compared with  mTPI-2, R-TPI, TITE-CRM and TITE-BOIN. 
Values shown are averages over simulated trials and the 60 dose-toxicity scenarios.
The unit of PCS, PCA, POA, POS and POT is \%, and the unit of duration (Dur) is day.} 
%\textperthousand, 
\label{tbl:simu_result}
\begin{center}
\scalebox{0.9}{
\begin{tabular}{c|rrrrrr}
\hline \hline
Method & PCS & PCA & POA & POS & POT & Dur \\ \hline
& \multicolumn{6}{c}{Setting 1} \\
mTPI-2 & 52.3 & 38.2 & 24.9 & 17.5 & 16.3 & 458 \\
R-TPI & 51.0 & \textbf{38.9} & \textbf{22.8} & \textbf{15.6} & \textbf{15.7} & 400 \\
TITE-CRM & 51.2 & 38.4 & 24.3 & 28.1 & 16.3 & \textbf{280} \\
TITE-BOIN & 51.7 & 37.3 & 25.2 & 24.8 & 16.3 & 336 \\
PoD-TPI & \textbf{52.2} & 38.2 & 24.0 & 17.4 & 16.1 & 389 \\ \hline
& \multicolumn{6}{c}{Setting 2} \\
mTPI-2 & 52.5 & 38.0 & 25.8 & 17.9 & 16.4 & 339 \\
R-TPI & 51.4 & \textbf{38.4} & 22.9 & \textbf{15.8} & 15.8 & 300 \\
TITE-CRM & 48.6 & 34.9 & \textbf{18.2} & 29.7 & \textbf{14.4} & \textbf{155} \\
TITE-BOIN & 51.4 & 37.1 & 24.6 & 23.8 & 16.1 & 227 \\
PoD-TPI & \textbf{51.5} & 37.6 & 23.2 & 17.2 & 15.8 & 265 \\ \hline
& \multicolumn{6}{c}{Setting 3} \\
mTPI-2 & 53.1 & 38.3 & 25.3 & 17.5 & 16.3 & 469 \\
R-TPI & 52.0 & \textbf{38.8} & 24.6 & \textbf{16.6} & 16.3 & 413 \\
TITE-CRM & \textbf{52.4} & 37.6 & 27.0 & 27.3 & 17.0 & \textbf{282} \\
TITE-BOIN & 52.0 & 37.0 & 27.4 & 26.2 & 17.0 & 343 \\
PoD-TPI & 52.3 & 37.9 & \textbf{24.4} & 17.8 & \textbf{16.2} & 402 \\ \hline
& \multicolumn{6}{c}{Setting 4} \\
mTPI-2 & 52.7 & 38.1 & 25.9 & 17.5 & 16.5 & 351 \\
R-TPI & 51.8 & \textbf{38.1} & 24.5 & \textbf{17.2} & 16.2 & 315 \\
TITE-CRM & 49.6 & 34.0 & \textbf{22.7} & 29.6 & \textbf{15.5} & \textbf{156} \\
TITE-BOIN & \textbf{52.1} & 36.7 & 27.5 & 26.1 & 16.9 & 234 \\
POD-TPI & 51.5 & 37.6 & 24.0 & 17.5 & 16.1 & 279 \\
\hline \hline
\end{tabular}
}
\end{center}
\end{table}

The frequencies of inconsistent decisions are reported in Table \ref{tbl:simu_inconsist_dec}, where 
the dose assignment decisions of the TITE designs are compared with the decisions that would be made by mTPI-2 if complete outcomes were observed.
An alternative way of defining inconsistent decisions is to compare each TITE decision with its complete data counterpart, e.g., comparing TITE-CRM with CRM and comparing TITE-BOIN with BOIN.
The frequencies of inconsistent decisions defined in this way is reported in Supplementary Table \ref{supp:tbl:inconsist1}, which is similar to those reported in Table \ref{tbl:simu_inconsist_dec}.
As discussed in Section \ref{sec:trial_example}, the chance of making inconsistent decisions is a major concern for drug companies and regulatory agencies. It is impossible to eliminate inconsistent decisions in the presence of pending outcomes, but the chance of making such decisions can be controlled by suspension rules.
We report a few summaries of Table \ref{tbl:simu_inconsist_dec}.

\begin{enumerate}
[noitemsep,nolistsep,leftmargin=*]
\item By applying confidence thresholds to decisions, PoD-TPI has the smallest frequencies of risky inconsistent decisions (DS, DE and SE) among all designs.
\item In particular, PoD-TPI never makes risky escalations (DE and SE). 
\item Notably, on average, the frequency of overly conservative inconsistent decisions made by PoD-TPI is also the smallest among all designs.
\end{enumerate}
We note that the inconsistent decisions do not necessarily affect the metrics like PCS, PCA, POA, POS and POT, since
the selection of MTD is always based on complete data.
Also, the inconsistent decisions can be either too aggressive or too conservative thus cancel out in the calculation of PCA and POA.
Instead, the inconsistent decisions are more about patients and their risks of being exposed to toxic doses.
If a dose escalation decision has large uncertainty, it is unethical to expose future patients to higher doses, and the safety review boards should express concerns regarding such decision.
From this point of view, PoD-TPI is more favorable compared to the other TITE designs.

\begin{table}[ht!]
\caption{Frequencies of inconsistent decisions compared to the decisions that would be made by mTPI-2 if complete outcomes were observed. Values are shown in $1/10^{3}$. The decisions in square brackets are inconsistent and risky.} 
%\textperthousand, 
\label{tbl:simu_inconsist_dec}
\begin{center}
\scalebox{0.9}{
\begin{tabular}{c|R{2.3em}R{2.3em}R{2.3em}R{2.3em}R{2.3em}R{2.3em}|R{2.3em}}
\hline \hline
Method & [DS] & [DE] & [SE] & SD & ED & ES & Sum \\ \hline
& \multicolumn{6}{c|}{Setting 1} & \\
R-TPI & 30.7 & \textbf{0.0} & \textbf{0.0} & \textbf{0.0} & \textbf{0.0} & 93.7 & 124.4 \\
TITE-CRM & 99.3 & 4.1 & 14.6 & 4.8 & 1.5 & 108.1 & 232.4 \\
TITE-BOIN & 10.9 & 5.8 & 30.4 & 46.6 & 0.2 & 20.4 & 114.3 \\ 
PoD-TPI & \textbf{6.1} & \textbf{0.0} & \textbf{0.0} & 25.0 & 1.0 & \textbf{7.9}  & \textbf{40.0} \\ \hline
& \multicolumn{6}{c|}{Setting 2} & \\
R-TPI & 40.3 & \textbf{0.0} & \textbf{0.0} & \textbf{0.0} & \textbf{0.0} & 84.3 & 124.6 \\
TITE-CRM & 66.2 & 3.8 & 14.7 & 2.7 & 1.0 & 69.8 & 158.2 \\
TITE-BOIN & 15.3 & 6.7 & 26.8 & 51.5 & 0.6 & 33.7 & 134.6\\
PoD-TPI & \textbf{7.8} & \textbf{0.0} & \textbf{0.0} & 43.9 & 0.3 & \textbf{20.9} & \textbf{72.9} \\ \hline
& \multicolumn{6}{c|}{Setting 3} & \\
R-TPI & 49.3 & \textbf{0.0} & \textbf{0.0} & \textbf{0.0} & \textbf{0.0} & 94.4 & 143.7 \\
TITE-CRM & 103.6 & 7.8 & 18.3 & 6.1 & 2.9 & 92.6 & 231.3 \\
TITE-BOIN & 22.9 & 11.3 & 40.0 & 39.8 & 0.1 & 17.5 & 131.6 \\
PoD-TPI & \textbf{11.7} & \textbf{0.0} & \textbf{0.0} & 23.1 & 1.1 & \textbf{9.8} & \textbf{45.7} \\ \hline
& \multicolumn{6}{c|}{Setting 4} & \\
R-TPI & 57.5 & \textbf{0.0} & \textbf{0.0} & \textbf{0.0} & \textbf{0.0} & 89.9 & 147.4 \\
TITE-CRM & 53.3 & 7.2 & 19.8 & 3.1 & 1.7 & 50.9 & 136.0 \\
TITE-BOIN & 31.1 & 12.1 & 37.8 & 41.4 & 0.4 & 28.1 & 150.9 \\
POD-TPI & \textbf{16.7} & \textbf{0.0} & \textbf{0.0} & 35.9 & 0.4 & \textbf{23.6} & \textbf{76.6} \\
\hline \hline
\end{tabular}
}
\end{center}
\end{table}

\paragraph{Sensitivity analysis.}

An attractive feature of PoD-TPI is the flexibility in calibrating the values of $\pi_{\E}$ and $\pi_{\text{D}}$ to achieve the desired trade-off between speed and safety.
The threshold $\pi_{\E}$ controls the rates of DE and SE decisions, and $\pi_{\text{D}}$ controls the rate of SD decisions. They both affect the speed of the trial through suspension rules.

We conduct additional simulations to demonstrate the performance of PoD-TPI with five combinations of $\pi_{\E}$ and $\pi_{\text{D}}$. The results are summarized in Table \ref{tbl:simu_result_SA}.
By setting $\pi_{\E} = 1$ and $\pi_{\text{D}} = 0$, we achieve the safest version of PoD-TPI, which never makes any inconsistent and risky decision (DS, DE and SE) and has the highest PCA and the lowest POA. As a  trade-off, the trial duration of the safest version is longer than that of the default version by 20 days.
On the other hand, by setting $\pi_{\E} = 0.8$ and $\pi_{\text{D}} = 0.25$, we obtain the fastest version of PoD-TPI, the speed of which is comparable to TITE-BOIN. The trial duration of the fastest version is shorter than that of the default version by 50 days. As a trade-off, the safety profile of the fastest version is worse than the other versions.

\begin{table}[ht!]
\caption{Performance of PoD-TPI with different choices of $\pi_{\E}$ and $\pi_{\text{D}}$.
Values shown are averages over simulated trials and the 60 dose-toxicity scenarios.
The unit of PCS, PCA and POA is \%, the unit of inconsistent decisions (DS, DE, SE, SD, ED, ES) is $1/10^3$, and the unit of duration (Dur) is day. The decisions in square brackets are inconsistent and risky.} 
%\textperthousand, 
\label{tbl:simu_result_SA}
\begin{center}
\scalebox{0.8}{
\begin{tabular}{cc|rrrrrrrrrr}
\hline \hline
$\pi_{\E}$ & $\pi_{\text{D}}$ & PCS & PCA & POA & Dur & [DS] & [DE] & [SE] & SD & ED & ES \\ \hline
& &  \multicolumn{10}{c}{Setting 1} \\
\multicolumn{2}{c|}{(mTPI-2)} & 52.3 & 38.2 & 24.9 & 458 & 0.0 & 0.0 & 0.0 & 0.0 & 0.0 & 0.0 \\
1 & 0.15 & \textbf{52.2} & 38.2 & 24.0 & 389 & 6.1 & \textbf{0.0} & \textbf{0.0} & 25.0 & \textbf{1.0} & 7.9 \\
1 & 0 & 51.8 & \textbf{38.2} & \textbf{23.8} & 410 & \textbf{0.0} & \textbf{0.0} & \textbf{0.0} & \textbf{24.9} & 1.1 & \textbf{7.3} \\
0.9 & 0.25 & 51.9 & 38.0 & 24.5 & 352 & 11.5 & 1.5 & 5.0 & 25.5 & 1.2 & 11.5 \\
0.9 & 0.15 & 52.0 & 37.9 & 24.5 & 359 & 6.1 & 1.5 & 5.1 & 25.5 & \textbf{1.0} & 7.9 \\
0.8 & 0.25 & \textbf{52.2} & 37.9 & 25.2 & \textbf{336} & 11.8 & 4.0 & 11.2 & 25.7 & 1.2 & 11.1 \\ \hline
& &  \multicolumn{10}{c}{Setting 2} \\
\multicolumn{2}{c|}{(mTPI-2)} & 52.5 & 38.0 & 25.8 & 339 & 0.0 & 0.0 & 0.0 & 0.0 & 0.0 & 0.0 \\
1 & 0.15 & 51.5 & 37.6 & 23.2 & 265 & 7.8 & \textbf{0.0} & \textbf{0.0} & 43.9 & \textbf{0.3} & 20.9 \\
1 & 0 & 51.5 & \textbf{37.7} & \textbf{22.9} & 285 & \textbf{0.0} & \textbf{0.0} & \textbf{0.0} & \textbf{43.0} & \textbf{0.3} & \textbf{19.4} \\
0.9 & 0.25 & 51.4 & 37.3 & 23.9 & 231 & 13.6 & 1.9 & 6.6 & 44.4 & 1.5 & 28.5 \\
0.9 & 0.15 & \textbf{51.7} & 37.5 & 24.0 & 237 & 7.8 & 1.9 & 6.7 & 45.1 & 1.1 & 20.0 \\
0.8 & 0.25 & 51.2 & 36.9 & 24.5 & \textbf{221} & 13.7 & 4.2 & 13.0 & 44.6 & 1.6 & 27.7 \\ \hline
& &  \multicolumn{10}{c}{Setting 3} \\
\multicolumn{2}{c|}{(mTPI-2)} & 53.1 & 38.3 & 25.3 & 469 & 0.0 & 0.0 & 0.0 & 0.0 & 0.0 & 0.0 \\
1 & 0.15 & 52.3 & 37.9 & 24.4 & 402 & 11.7 & \textbf{0.0} & \textbf{0.0} & 23.1 & \textbf{1.1} & 9.8 \\
1 & 0 & 52.2 & \textbf{38.1} & \textbf{24.0} & 422 & \textbf{0.0} & \textbf{0.0} & \textbf{0.0} & \textbf{22.8} & \textbf{1.1} & \textbf{8.9} \\
0.9 & 0.25 & \textbf{52.5} & 37.5 & 26.0 & 367 & 21.8 & 4.4 & 10.7 & 23.3 & 1.3 & 14.3 \\
0.9 & 0.15 & 52.4 & 37.7 & 25.8 & 374 & 12.6 & 4.5 & 10.9 & 23.9 & \textbf{1.1} & 9.2 \\
0.8 & 0.25 & 52.3 & 37.4 & 27.5 & \textbf{350} & 22.6 & 9.8 & 21.3 & 24.2 & 1.2 & 13.1 \\ \hline
& &  \multicolumn{10}{c}{Setting 4} \\
\multicolumn{2}{c|}{(mTPI-2)} & 52.7 & 38.1 & 25.9  & 351 & 0.0 & 0.0 & 0.0 & 0.0 & 0.0 & 0.0 \\
1 & 0.15 & 51.7 & 37.6 & 23.9 & 279 & 16.9 & \textbf{0.0} & \textbf{0.0} & 36.3 & 0.4 & 23.9 \\
1 & 0 & 51.5 & \textbf{37.7} & \textbf{23.2} & 299 & \textbf{0.0} & \textbf{0.0} & \textbf{0.0} & \textbf{36.0} & \textbf{0.3} & \textbf{21.7} \\
0.9 & 0.25 & \textbf{52.1} & 36.9 & 25.9 & 246 & 28.9 & 6.1 & 14.5 & 37.0 & 1.3 & 31.1 \\
0.9 & 0.15 & 52.0 & 37.2 & 26.0 & 253 & 18.1 & 6.1 & 15.3 & 38.0 & 0.7 & \textbf{21.7} \\
0.8 & 0.25 & \textbf{52.1} & 36.5 & 27.6 & \textbf{236} & 29.7 & 11.7 & 25.9 & 38.0 & 1.4 & 29.1 \\
\hline \hline
\end{tabular}
}
\end{center}
\end{table}

\paragraph{Benchmark against rolling six.}
The rolling six (R6) design \citep{skolnik2008shortening} is a widely used phase I design that allows for continual accrual of up to six patients onto a dose level. 
So far, it has 139 citations (according to Google Scholar) and was used in many real-world trials (e.g., \citealp{mosse2013safety} and \citealp{von2016phase}).
R6 consists of a set of algorithmic decision rules that determine the dose-finding decisions and MTD selection. 
The rules are transparent and are interpretable to clinicians. 
%R6 targets a DLT rate of 0.17 and does not allow for other targets; it also does not allow for a pre-specified maximum sample size.

We conduct additional simulations to benchmark the performance of PoD-TPI against R6. 
The aim is to show that PoD-TPI has a lower risk than R6, thereby further establish the potential utility of the PoD-TPI design in practice.
For a fair comparison, we consider the 20 dose-toxicity scenarios with a target DLT probability of $p_{\T} = 0.17$ and calibrate the maximum sample size of PoD-TPI to match the average sample size of R6. Specifically, for each scenario, we first simulate 1,000 trials using R6 as the design. Next, we calculate the average sample size for R6 across the simulated trials. Finally, we round up the average sample size for R6 and use it as the maximum sample size for PoD-TPI.
The average sample size of R6 across the 20 scenarios is around 19 patients.

The simulation results of R6 and PoD-TPI are summarized in Tables \ref{tbl:simu_result_r6} and \ref{tbl:simu_inconsist_dec_r6}. The trial speed of R6 is faster than that of PoD-TPI. However, the reliability and safety of PoD-TPI are uniformly better than R6. In practice, usually safety is the most important criterion, thus PoD-TPI is more favorable compared to R6.

\begin{table}[ht!]
\caption{Performance of PoD-TPI compared with  rolling six (R6). 
Values shown are averages over simulated trials and the 60 dose-toxicity scenarios.
The unit of PCS, PCA, POA, POS and POT is \%, and the unit of duration (Dur) is day.} 
%\textperthousand, 
\label{tbl:simu_result_r6}
\begin{center}
\scalebox{0.9}{
\begin{tabular}{c|rrrrrr}
\hline \hline
Method & PCS & PCA & POA & POS & POT & Dur \\ \hline
& \multicolumn{6}{c}{Setting 1} \\
R6 & 48.4 & 34.2 & 24.9 & 27.7 & 17.5 & \textbf{245} \\
PoD-TPI & \textbf{50.5} & \textbf{37.6} & \textbf{19.4} & \textbf{14.8} & \textbf{13.5} & 299 \\ \hline
& \multicolumn{6}{c}{Setting 2} \\
R6 & 48.7 & 34.0 & 25.4 & 27.6 & 17.3 & \textbf{175} \\
PoD-TPI & \textbf{49.0} & \textbf{36.9} & \textbf{19.6} & \textbf{15.7} & \textbf{13.4} & 217 \\\hline
& \multicolumn{6}{c}{Setting 3} \\
R6 & 48.4 & 34.1 & 25.4 & 27.6 & 17.4 & \textbf{251} \\
PoD-TPI & \textbf{49.8} & \textbf{37.2} & \textbf{19.7} & \textbf{16.2} & \textbf{13.4} & 311 \\
\hline
& \multicolumn{6}{c}{Setting 4} \\
R6 & 48.3 & 33.7 & 25.7 & 27.5 & 17.3 & \textbf{177} \\
POD-TPI & \textbf{49.6} & \textbf{36.9} & \textbf{19.8} & \textbf{15.4} & \textbf{13.4} & 225 \\
\hline \hline
\end{tabular}
}
\end{center}
\end{table}

\begin{table}[h!]
\caption{Frequencies of inconsistent decisions compared to the decisions that would be made by mTPI-2 if complete outcomes were observed. Values are shown in $1/10^{3}$. The decisions in square brackets are inconsistent and risky.} 
%\textperthousand, 
\label{tbl:simu_inconsist_dec_r6}
\begin{center}
\scalebox{0.9}{
\begin{tabular}{c|R{2.3em}R{2.3em}R{2.3em}R{2.3em}R{2.3em}R{2.3em}|R{2.3em}}
\hline \hline
Method & [DS] & [DE] & [SE] & SD & ED & ES & Sum \\ \hline
& \multicolumn{6}{c|}{Setting 1} & \\
R6 & 118.1 & \textbf{0.0} & 43.0 & \textbf{0.0} & \textbf{0.0} & 183.7 & 344.8 \\
PoD-TPI & \textbf{4.5} & \textbf{0.0} & \textbf{0.0} & 17.8 & 1.6 & \textbf{3.5} & \textbf{27.4} \\  \hline
& \multicolumn{6}{c|}{Setting 2} & \\
R6 & 77.9 & \textbf{0.0} & 43.7 & \textbf{0.0} & \textbf{0.0} & 51.1 & 172.7 \\
PoD-TPI & \textbf{6.7} & \textbf{0.0} & \textbf{0.0} & 37.9 & 0.5 & \textbf{10.7} & \textbf{55.8} \\  \hline
& \multicolumn{6}{c|}{Setting 3} & \\
R6 & 95.4 & \textbf{0.0} & 42.3 & \textbf{0.0} & \textbf{0.0} & 180.9 & 318.6 \\
PoD-TPI & \textbf{9.6} & \textbf{0.0} & \textbf{0.0} & 22.8 & 1.8 & \textbf{3.9} & \textbf{38.1} \\
\hline
& \multicolumn{6}{c|}{Setting 4} & \\
R6 & 32.5 & \textbf{0.0} & 43.0 & \textbf{0.0} & \textbf{0.0} & 50.4 & 125.9 \\
POD-TPI & \textbf{14.4} & \textbf{0.0} & \textbf{0.0} & 43.2 & 0.5 & \textbf{13.2} & \textbf{71.3} \\
\hline \hline
\end{tabular}
}
\end{center}
\end{table}

\section{Discussion}
\label{sec:discussion}

We have proposed the PoD-TPI design to accelerate phase I trials. 
We have developed a statistical methodology to calculate the posterior distribution of a dose assignment decision in the presence of pending toxicity outcomes.
The posterior distribution directly reflects the confidence of all possible decisions, and the optimal decision is computed under a decision-theoretic framework. Intuitive suspension rules are added based on the risk of the optimal decision, minimizing the chance of making inconsistent decisions.
Specifically, the probability threshold for suspension can be flexibly adjusted to balance between speed and safety.

The PoD-TPI design is built upon the mTPI-2 design. Nevertheless, the proposed strategy can be applied to other model-free or model-assisted designs such as 3+3 \citep{storer1989design}, BOIN \citep{liu2015bayesian}, keyboard \citep{yan2017keyboard} or i3+3 \citep{liu2019design}. It suffices to change the deterministic decision function $\A$ according to the specific design.

The PoD-TPI design has an early stopping rule when the lowest dose is too toxic. If desired, another early stopping rule can be added when the MTD has been identified with high confidence. For example, when $\Pr(p_d \in I_{\E} \mid \data) > 0.95$ for some dose level $d$.

In PoD-TPI, the likelihood of the unknown parameters $\bp$ and $\bw$ is constructed based on survival modeling (Equation \ref{eq:likelihood}), which implicitly assumes that all of the study subjects will eventually experience the event of interest (DLT) if they are followed long enough. Such assumption may not be valid for immunotherapies, because they may not induce DLT events at all in some patients. In such cases, we can use a cure rate model \citep{boag1949maximum}, assuming with non-zero probability that patients may never experience DLT. Little would change in the overall setup; see Supplementary Section \ref{supp:sec:cure_rate} for a discussion.

%To further speed up the trial, it is possible to relax the trial suspension criteria in PoD-TPI at the cost of more inconsistent decisions. In the extreme case, we can discard the suspension rule to achieve the optimal speed.
%An alternative strategy is to be more conservative, assigning more patients to lower doses (instead of suspending the trial) when we are uncertain about an escalation decision. The drawback of this strategy is the possibility of treating more patients at subtherapeutic doses.

Finally, in practice, it is desirable to tabulate the dose decision rules of the design before the trial begins, so that the rules are transparent and can be examined by clinicians.
We note that this is generally not possible for model-based or model-assisted time-to-event designs without some approximation (see, e.g., \citealp{yuan2018time} and \citealp{lin2018time}).
A future direction is to explore what approximation is needed for PoD-TPI so that its decision rules can be pre-tabulated.

\bibliographystyle{apalike}
\bibliography{PODTPI-ref}

\clearpage

\section*{Supplementary Materials}

\beginsupplement

\subsection{Probability of Decisions}
\label{supp:sec:pod}

We show that the PoD
\begin{align*}
\Pr(A_d = a \mid \data) = \sum_{s:  \A(n_d + s, m_d + r_d - s) = a} \Pr(S_d = s \mid \data)
\end{align*}
sums up to 1 over the space of $a \in \{-1, 0, 1\}$.
Note that for any $s \in \{ 0, 1, \ldots, r_d \}$, $\A(n_d + s, m_d + r_d - s) \in \{-1, 0, 1\}$. 
On the other hand, the support of $S_d$ is  $\{ 0, 1, \ldots, r_d \}$.
Therefore, 
\begin{align*}
\Pr \left(A_d \in \{-1, 0, 1\} \mid \data \right) = \sum_{s \in \{ 0, 1, \ldots, r_d \}} \Pr(S_d = s \mid \data) = 1.
\end{align*}

\subsection{Selection of the MTD}
\label{supp:sec:sel_mtd}

The trial is completed if the number of enrolled patients reaches the pre-specified maximum allowable sample size $N^*$ or  safety rule 1 is triggered.
The last step is to recommend an MTD based on the DLT outcomes.
If the trial is terminated due to safety rule 1, no MTD is selected. Otherwise, after DLT assessment for all patients is finished, we use the same procedure as in \cite{ji2007dose} to report an MTD. 
The isotonically transformed posterior means are used as estimates of $p_d$'s subject to the order constraint $p_1 \leq \cdots \leq p_D$.
In particular, we first calculate posterior means and variances of the DLT probabilities, $\{ \tilde{p}_1, \ldots, \tilde{p}_D \}$ and $\{ \nu_1, \ldots, \nu_D \}$.
Then, we solve the optimization problem, minimizing $\sum_{d = 1}^D  (\hat{p}_d - \tilde{p}_d)^2 / \nu_d$ subject to $\hat{p}_j \geq \hat{p}_i$ for $j > i$.
Such optimization can be done using the pooled adjacent violators algorithm, and $\{ \hat{p}_1, \ldots, \hat{p}_D \}$ are estimated DLT probabilities satisfying the order constraint.
Let $\D = \{ d: \hat{p}_d \in I_{\E} \}$ and $\D_{-} = \{ d: \hat{p}_d \in I_{\U} \}$.
The procedure of selecting the MTD is summarized in Algorithm \ref{alg:sel-mtd}.
We note that the recommended dose for the last patient needs not be the selected MTD.

\begin{algorithm}[h!]
\caption{MTD selection rule. Here $|\D|$ is the cardinality of set $\D$.}
\label{alg:sel-mtd}
\begin{algorithmic}[1]

\If{$|\D| = 0$}
\If{$|\D_{-}| \geq 1$}
\State Select the highest dose in $\D_{-}$ as the MTD
\Else
\State No MTD is selected
\EndIf
\ElsIf{$|\D| = 1$}
\State Select the dose in $\D$ as the MTD
\ElsIf{$|\D| > 1$}
\State Let $\D_0 = \argmin_{d \in \D} |\hat{p}_d - p_{\T}|$
\If{$|\D_0| = 1$}
\State Select the dose in $\D_0$ as the MTD
\ElsIf{$|\D_0| > 1$}
\State Let $\D_{0-} = \{ d: \hat{p}_d \in \D_0,  \hat{p}_d \in [p_{\T} - \epsilon_1, p_{\T}] \}$
\If{$|\D_{0-}| \geq 1$}
\State Select the highest dose in $\D_{0-}$ as the MTD
\Else
\State Select the lowest dose in $\D_{0}$ as the MTD
\EndIf
\EndIf
\EndIf
\end{algorithmic}
\end{algorithm}

\clearpage

\subsection{Simulation Details}

\begin{figure}[h!]
\begin{center}
\includegraphics[width=.95\textwidth]{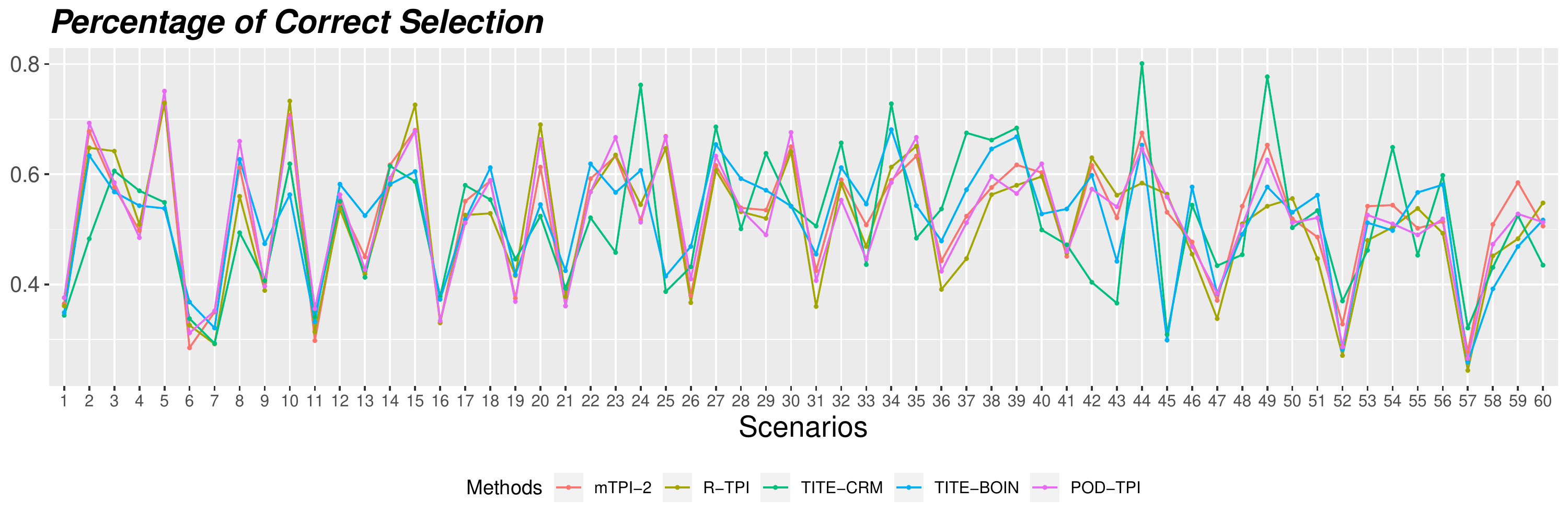}
\includegraphics[width=.95\textwidth]{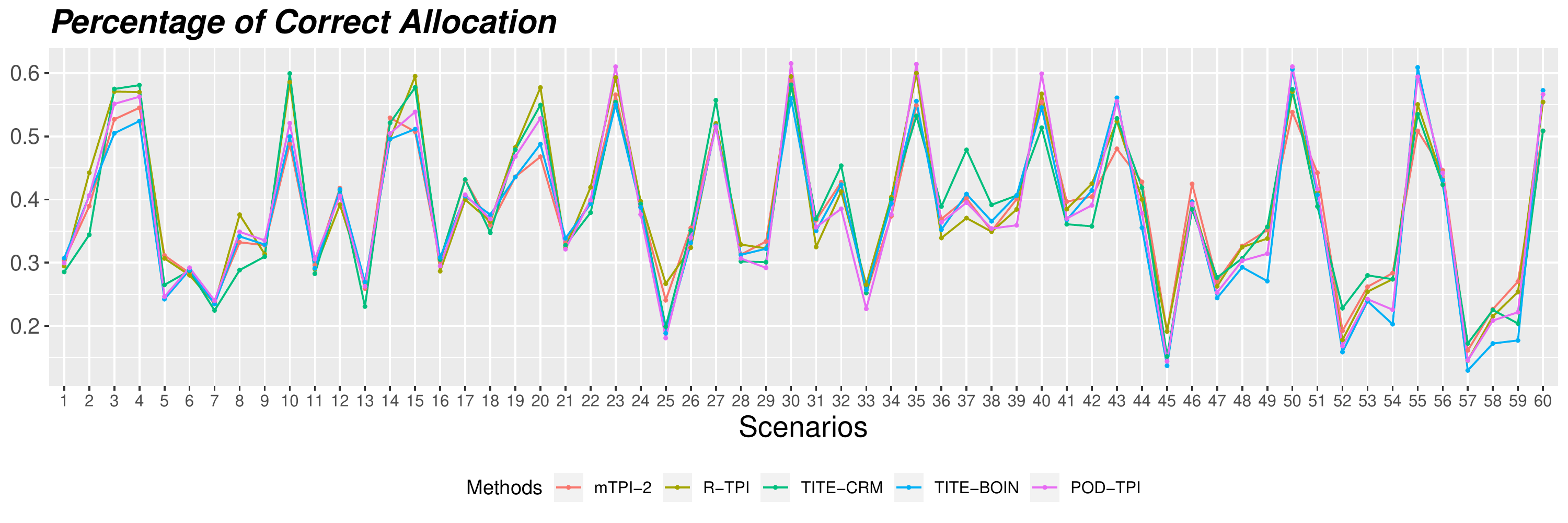}
\includegraphics[width=.95\textwidth]{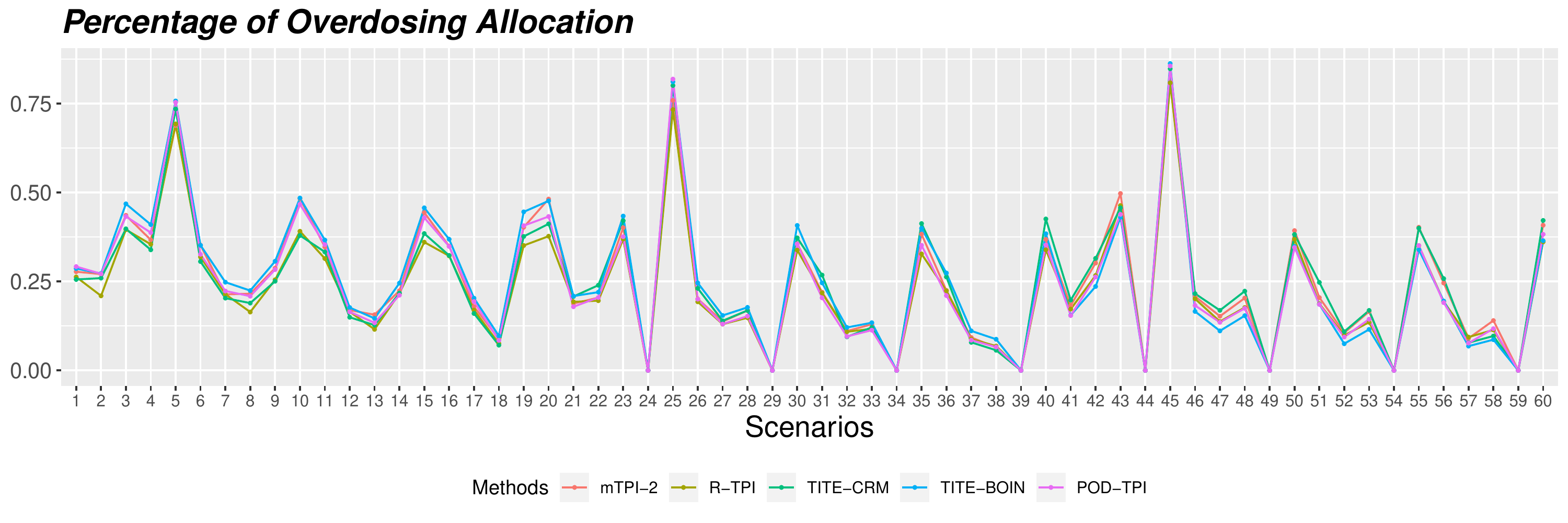}
\end{center}
\caption{Scenario-specific PCS, PCA and POA for mTPI-2, R-TPI, TITE-CRM, TITE-BOIN and PoD-TPI under Setting 1. }
\label{supp:fig:pcs}
\end{figure}

\begin{figure}[h!]
\begin{center}
\includegraphics[width=.95\textwidth]{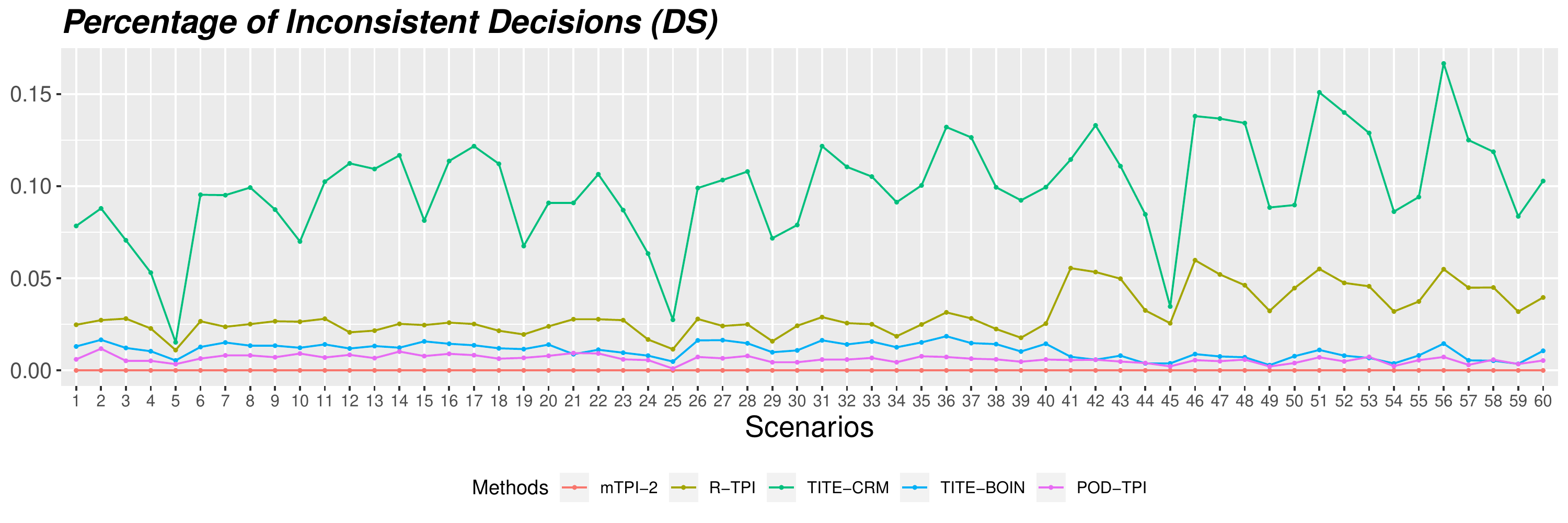}
\includegraphics[width=.95\textwidth]{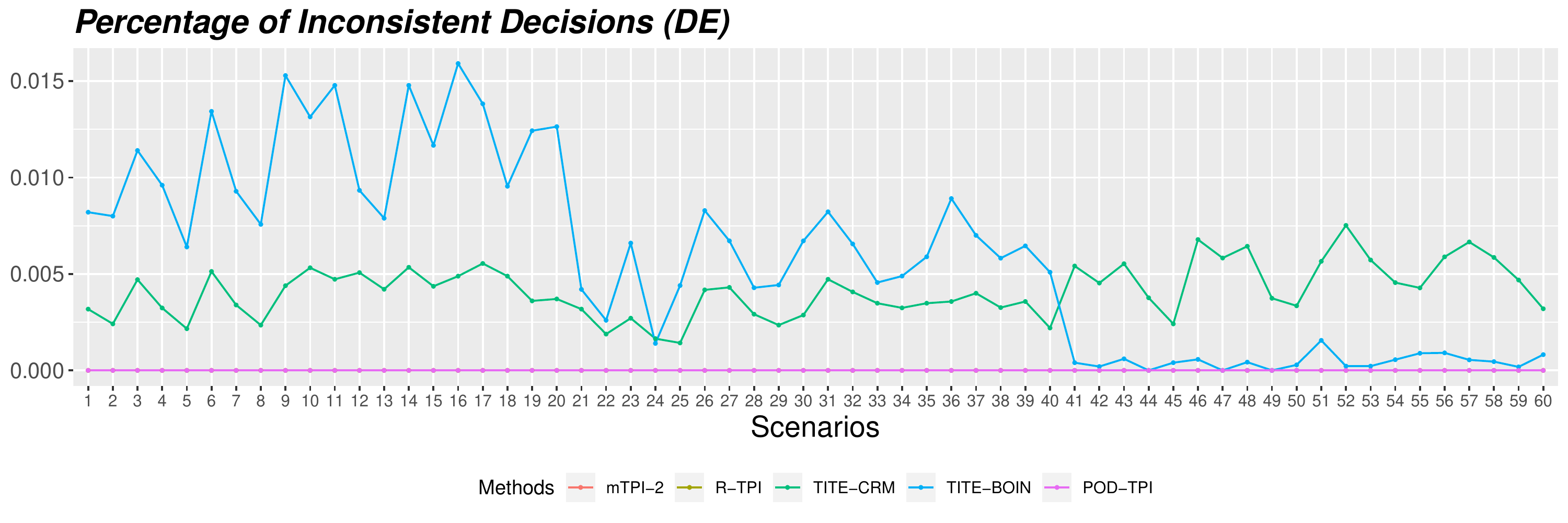}
\includegraphics[width=.95\textwidth]{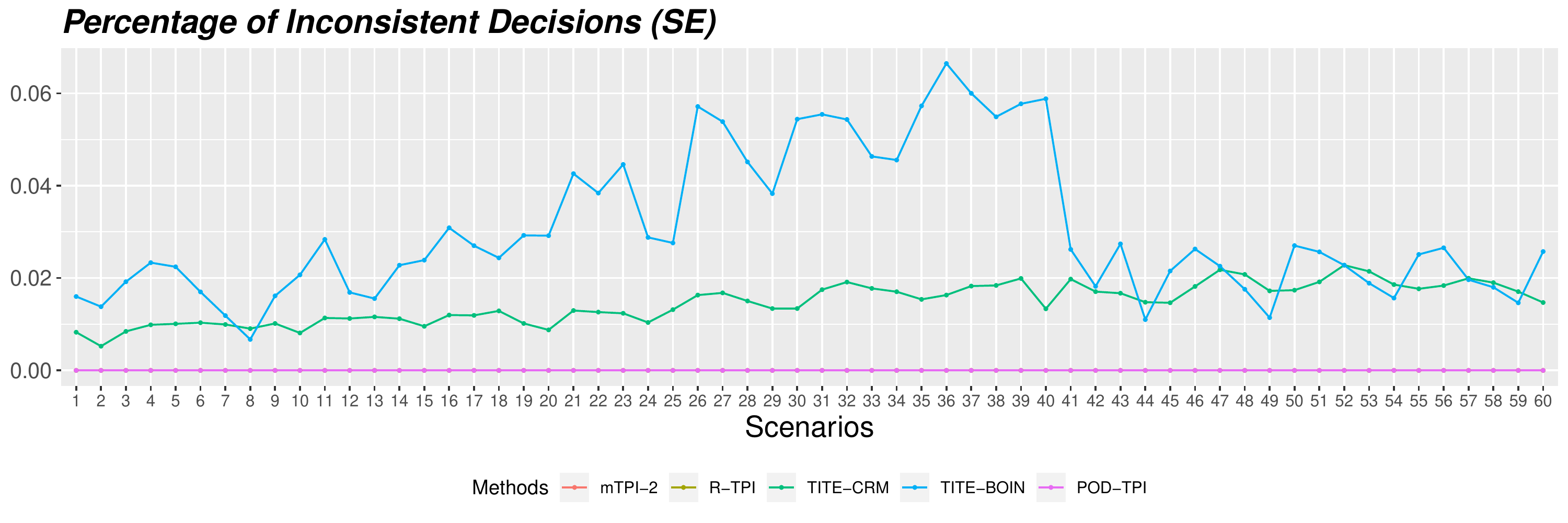}
\end{center}
\caption{Scenario-specific frequencies of inconsistent and risky decisions for mTPI-2, R-TPI, TITE-CRM, TITE-BOIN and PoD-TPI under Setting 1. }
\label{supp:fig:inconsist}
\end{figure}

We provide supplementary tables and figures that are referenced in the manuscript.
Table \ref{tbl:simu_truth} summarizes the 60 dose-toxicity scenarios used in the simulation studies. 
Tables \ref{supp:tbl:inconsist1} and \ref{supp:tbl:inconsist2} report the frequencies of inconsistent decisions by comparing the decisions made by R-TPI, TITE-CRM, TITE-BOIN and PoD-TPI with their complete data counterparts.
Figures \ref{supp:fig:pcs} and \ref{supp:fig:inconsist} show the scenario-specific operating characteristics.

\setlength{\LTcapwidth}{\textwidth}
\begin{longtable}{|c|c|c|rrrrrr|}
\caption{60 dose-toxicity scenarios with different target DLT probabilities $p_{\T} \in \{ 0.10, 0.17, 0.30 \}$, numbers of doses $D \in \{ 3, 4, 5, 6\}$ and true DLT probabilities $\{ p_1, \ldots, p_D \}$. } 
\label{tbl:simu_truth} \\
%This part appears at the top of the table
\hline
Scn. & $p_{\T}$ & $D$ & $p_1$ & $p_2$ & $p_3$ & $p_4$ & $p_5$ & $p_6$ \\ \hline
\endfirsthead
\caption[]{(continued)}\\
\hline
Scn. & $p_{\T}$ & $D$ & $p_1$ & $p_2$ & $p_3$ & $p_4$ & $p_5$ & $p_6$ \\ \hline
\endhead
% the end of the table (each page)
\hline
\endfoot
% the total end of the table
\hline
\endlastfoot
1 & 0.10 & 3 & 0.05 & 0.10 & 0.15 &  &  &  \\
2 & 0.10 & 3 & 0.03 & 0.06 & 0.28 &  &  &  \\
3 & 0.10 & 3 & 0.06 & 0.20 & 0.30 &  &  &  \\
4 & 0.10 & 3 & 0.10 & 0.20 & 0.30 &  &  &  \\
5 & 0.10 & 3 & 0.21 & 0.32 & 0.43 &  &  &  \\
6 & 0.10 & 4 & 0.05 & 0.10 & 0.15 & 0.20 &  &  \\
7 & 0.10 & 4 & 0.03 & 0.06 & 0.10 & 0.15 &  &  \\
8 & 0.10 & 4 & 0.02 & 0.04 & 0.06 & 0.28 &  &  \\
9 & 0.10 & 4 & 0.05 & 0.10 & 0.20 & 0.30 &  &  \\
10 & 0.10 & 4 & 0.05 & 0.20 & 0.35 & 0.50 &  &  \\
11 & 0.10 & 5 & 0.05 & 0.10 & 0.15 & 0.20 & 0.25 &  \\
12 & 0.10 & 5 & 0.02 & 0.04 & 0.08 & 0.10 & 0.16 &  \\
13 & 0.10 & 5 & 0.02 & 0.04 & 0.06 & 0.08 & 0.28 &  \\
14 & 0.10 & 5 & 0.03 & 0.07 & 0.12 & 0.15 & 0.25 &  \\
15 & 0.10 & 5 & 0.06 & 0.19 & 0.32 & 0.45 & 0.58 &  \\
16 & 0.10 & 6 & 0.05 & 0.10 & 0.15 & 0.20 & 0.25 & 0.30 \\
17 & 0.10 & 6 & 0.02 & 0.05 & 0.08 & 0.12 & 0.16 & 0.20 \\
18 & 0.10 & 6 & 0.02 & 0.04 & 0.06 & 0.08 & 0.10 & 0.28 \\
19 & 0.10 & 6 & 0.11 & 0.15 & 0.21 & 0.25 & 0.30 & 0.35 \\
20 & 0.10 & 6 & 0.06 & 0.17 & 0.28 & 0.39 & 0.50 & 0.61 \\
21 & 0.17 & 3 & 0.08 & 0.17 & 0.25 &  &  &  \\
22 & 0.17 & 3 & 0.06 & 0.12 & 0.34 &  &  &  \\
23 & 0.17 & 3 & 0.10 & 0.26 & 0.35 &  &  &  \\
24 & 0.17 & 3 & 0.04 & 0.08 & 0.12 &  &  &  \\
25 & 0.17 & 3 & 0.27 & 0.37 & 0.47 &  &  &  \\
26 & 0.17 & 4 & 0.08 & 0.17 & 0.25 & 0.33 &  &  \\
27 & 0.17 & 4 & 0.06 & 0.12 & 0.17 & 0.23 &  &  \\
28 & 0.17 & 4 & 0.04 & 0.08 & 0.12 & 0.34 &  &  \\
29 & 0.17 & 4 & 0.03 & 0.06 & 0.09 & 0.12 &  &  \\
30 & 0.17 & 4 & 0.12 & 0.26 & 0.40 & 0.54 &  &  \\
31 & 0.17 & 5 & 0.08 & 0.17 & 0.25 & 0.33 & 0.41 &  \\
32 & 0.17 & 5 & 0.04 & 0.08 & 0.12 & 0.17 & 0.25 &  \\
33 & 0.17 & 5 & 0.03 & 0.06 & 0.09 & 0.12 & 0.34 &  \\
34 & 0.17 & 5 & 0.03 & 0.06 & 0.09 & 0.12 & 0.15 &  \\
35 & 0.17 & 5 & 0.13 & 0.25 & 0.37 & 0.49 & 0.61 &  \\
36 & 0.17 & 6 & 0.08 & 0.17 & 0.25 & 0.33 & 0.41 & 0.49 \\
37 & 0.17 & 6 & 0.03 & 0.10 & 0.15 & 0.20 & 0.25 & 0.30 \\
38 & 0.17 & 6 & 0.03 & 0.06 & 0.09 & 0.12 & 0.15 & 0.34 \\
39 & 0.17 & 6 & 0.04 & 0.08 & 0.10 & 0.12 & 0.14 & 0.16 \\
40 & 0.17 & 6 & 0.14 & 0.24 & 0.34 & 0.44 & 0.54 & 0.64 \\
41 & 0.30 & 3 & 0.15 & 0.30 & 0.45 &  &  &  \\
42 & 0.30 & 3 & 0.10 & 0.20 & 0.44 &  &  &  \\
43 & 0.30 & 3 & 0.18 & 0.38 & 0.46 &  &  &  \\
44 & 0.30 & 3 & 0.08 & 0.16 & 0.24 &  &  &  \\
45 & 0.30 & 3 & 0.39 & 0.48 & 0.57 &  &  &  \\
46 & 0.30 & 4 & 0.15 & 0.30 & 0.45 & 0.60 &  &  \\
47 & 0.30 & 4 & 0.10 & 0.20 & 0.30 & 0.40 &  &  \\
48 & 0.30 & 4 & 0.08 & 0.16 & 0.24 & 0.44 &  &  \\
49 & 0.30 & 4 & 0.06 & 0.12 & 0.18 & 0.24 &  &  \\
50 & 0.30 & 4 & 0.26 & 0.38 & 0.50 & 0.62 &  &  \\
51 & 0.30 & 5 & 0.15 & 0.30 & 0.45 & 0.60 & 0.75 &  \\
52 & 0.30 & 5 & 0.08 & 0.16 & 0.24 & 0.30 & 0.38 &  \\
53 & 0.30 & 5 & 0.06 & 0.12 & 0.18 & 0.24 & 0.44 &  \\
54 & 0.30 & 5 & 0.05 & 0.10 & 0.15 & 0.20 & 0.25 &  \\
55 & 0.30 & 5 & 0.27 & 0.37 & 0.47 & 0.57 & 0.67 &  \\
56 & 0.30 & 6 & 0.14 & 0.30 & 0.44 & 0.58 & 0.72 & 0.86 \\
57 & 0.30 & 6 & 0.06 & 0.12 & 0.18 & 0.24 & 0.30 & 0.36 \\
58 & 0.30 & 6 & 0.05 & 0.10 & 0.15 & 0.20 & 0.25 & 0.44 \\
59 & 0.30 & 6 & 0.04 & 0.08 & 0.12 & 0.16 & 0.20 & 0.24 \\
60 & 0.30 & 6 & 0.27 & 0.36 & 0.45 & 0.54 & 0.63 & 0.72 \\
\end{longtable}

\begin{table}[h!]
\caption{Frequencies of inconsistent decisions compared to the decisions that would be made by the corresponding complete data design. 
R-TPI is compared to mTPI-2, TITE-CRM is compared to CRM, TITE-BOIN is compared to BOIN, and PoD-TPI is compared to mTPI-2.
Values are shown in $1/10^{3}$.} 
%\textperthousand, 
\label{supp:tbl:inconsist1}
\begin{center}
\begin{tabular}{|c|R{2.3em}R{2.3em}R{2.3em}R{2.3em}R{2.3em}R{2.3em}|R{2.3em}|}
\hline
Method & DS & DE & SE & SD & ED & ES & Sum \\ \hline
& \multicolumn{6}{c|}{Setting 1} & \\
R-TPI & 30.7 & \textbf{0.0} & \textbf{0.0} & \textbf{0.0} & \textbf{0.0} & 93.7 & 124.4 \\
TITE-CRM & 29.9 & 1.1 & 8.5 & 5.5 & 0.0 & 53.6 & 98.6 \\
TITE-BOIN & 11.0 & 5.8 & 15.2 & 46.6 & 0.2 & 31.8 & 110.6 \\
PoD-TPI & \textbf{6.1} & \textbf{0.0} & \textbf{0.0} & 25.0 & 1.0 & \textbf{7.9} & \textbf{40.0} \\
\hline
& \multicolumn{6}{c|}{Setting 2} & \\
R-TPI & 40.3 & \textbf{0.0} & \textbf{0.0} & \textbf{0.0} & \textbf{0.0} & 84.3 & 124.6\\
TITE-CRM & 20.8 & 0.5 & 10.4 & 4.7 & 0.0 & 70.3 & 106.7 \\
TITE-BOIN & 15.3 & 6.7 & 17.9 & 51.4 & 0.7 & 50.4 & 142.4 \\
PoD-TPI & \textbf{7.8} & \textbf{0.0} & \textbf{0.0} & 43.9 & 0.3 & \textbf{20.9} & \textbf{72.9} \\
\hline
& \multicolumn{6}{c|}{Setting 3} & \\
R-TPI & 49.3 & \textbf{0.0} & \textbf{0.0} & \textbf{0.0} & \textbf{0.0} & 94.4 & 143.7 \\
TITE-CRM & 49.2 & 2.8 & 15.5 & 4.1 & 0.0 & 39.4  & 111.0 \\
TITE-BOIN & 22.9 & 11.3 & 25.9 & 39.8 & 0.1 & 27.7 & 127.7 \\
PoD-TPI & \textbf{11.7} & \textbf{0.0} & \textbf{0.0} & 23.1 & 1.1 & \textbf{9.8} & \textbf{45.7} \\ \hline
& \multicolumn{6}{c|}{Setting 4} & \\
R-TPI & 57.5 & \textbf{0.0} & \textbf{0.0} & \textbf{0.0} & \textbf{0.0} & 89.9 & 147.4 \\
TITE-CRM & 30.6 & 1.9 & 18.2 & 2.7 & 0.0 & 37.2 & 90.6 \\
TITE-BOIN & 31.2 & 12.1 & 28.9 & 41.4 & 0.4 & 43.0 & 157.0 \\
POD-TPI & \textbf{16.7} & \textbf{0.0} & \textbf{0.0} & 35.9 & 0.4 & \textbf{23.6} & \textbf{76.6} \\
\hline
\end{tabular}
\end{center}
\end{table}

\begin{table}[h!]
\caption{Frequencies of inconsistent decisions compared to the decisions that would be made by the corresponding complete data design. 
R6 is compared to complete-data R6, and PoD-TPI is compared to mTPI-2.
Values are shown in $1/10^{3}$.} 
%\textperthousand, 
\label{supp:tbl:inconsist2}
\begin{center}
\begin{tabular}{|c|R{2.3em}R{2.3em}R{2.3em}R{2.3em}R{2.3em}R{2.3em}|R{2.3em}|}
\hline
Method & DS & DE & SE & SD & ED & ES & Sum \\ \hline
& \multicolumn{6}{c|}{Setting 1} & \\
R6 & 41.8 & \textbf{0.0} & \textbf{0.0} & \textbf{0.0} & \textbf{0.0} & 146.1 & 187.9 \\
PoD-TPI & \textbf{4.5} & \textbf{0.0} & \textbf{0.0} & 17.8 & 1.6 & \textbf{3.5} & \textbf{27.4} \\
\hline
& \multicolumn{6}{c|}{Setting 2} & \\
R6 & 43.8 & \textbf{0.0} & \textbf{0.0} & \textbf{0.0} & \textbf{0.0} & 47.8 & 91.6 \\
PoD-TPI & \textbf{6.7} & \textbf{0.0} & \textbf{0.0} & 37.9 & 0.5 & \textbf{10.7} & \textbf{55.8} \\
\hline
& \multicolumn{6}{c|}{Setting 3} & \\
R6 & 48.1 & \textbf{0.0} & \textbf{0.0} & \textbf{0.0} & \textbf{0.0} & 143.9 & 192.0  \\
PoD-TPI & \textbf{9.6} & \textbf{0.0} & \textbf{0.0} & 22.8 & 1.8 & \textbf{3.9} & \textbf{38.1} \\
\hline
& \multicolumn{6}{c|}{Setting 4} & \\
R6 & 24.1 & \textbf{0.0} & \textbf{0.0} & \textbf{0.0} & \textbf{0.0} & 47.0 & \textbf{71.1} \\
POD-TPI & \textbf{14.4} & \textbf{0.0} & \textbf{0.0} & 43.2 & 0.5 & \textbf{13.2} & 71.3 \\
\hline
\end{tabular}
\end{center}
\end{table}

\clearpage

\clearpage

\subsection{Cure Rate Model}
\label{supp:sec:cure_rate}

To construct a cure rate model, we introduce an additional notation $G_i$, such that $G_i = 1$ (or 0) represents that patient $i$ experiences DLT eventually (or never experiences DLT during his/her lifetime). We model $G_i$ with a Bernoulli distribution, $G_i \mid Z_i = d \sim \text{Bernoulli}(\xi_d)$. It is easy to see that $\xi_d \geq p_d$, as $Y_i = 1$ always implies  $G_i = 1$.
We have $\Pr(Y_i = 1 \mid G_i = 1, Z_i = d) = p_d / \xi_d$.

Still, we assume a piecewise uniform distribution for $[T_i \mid Y_i = 1, G_i = 1, Z_i = d]$ on the interval $(0, \tau]$. The conditional density of $T_i$ is
\begin{align*}
f_{T \mid Y, G, Z} (t \mid Y_i = 1, G_i = 1, Z_i = d, \bw) =  w_k \cdot \frac{1}{h_k - h_{k-1}}, \quad \text{for} \; h_{k-1} < t \leq h_k.
\end{align*}
Next,
\begin{multline*}
f_{T \mid Z}(t \mid Z_i = d, p_d, \xi_d, \bw)  
= \sum_{y \in \{ 0, 1 \}} \sum_{g \in  \{ 0, 1 \}} \Big[ f_{T \mid Y, G, Z} (t \mid Y_i = y, G_i = g, Z_i = d, \bw) \times \\
\Pr(Y_i = y \mid G_i = g, Z_i = d, p_d, \xi_d) \Pr(G_i = g \mid Z_i = d, \xi_d) \Big] \\
= w_k \cdot \frac{1}{h_k - h_{k-1}} \cdot \frac{p_d}{\xi_d} \cdot \xi_d + 0 + 0 + 0 = p_d \cdot w_k \cdot \frac{1}{h_k - h_{k-1}}, \quad \text{for} \; h_{k-1} < t \leq h_k,
\end{multline*}
and 
\begin{align*}
S_{T \mid Z} (t \mid Z_i = d, p_d, \xi_d, \bw) =  1 - p_d \sum_{k = 1}^K w_k \beta(t, k), \quad \text{for} \; t \leq \tau.
\end{align*}
We can see the cure rate model does not change anything in the likelihood (7), i.e., $\xi_d$'s do not contribute to the likelihood. This is expected, as the data only provide information for $\xi_d$ up to $p_d$ ($\xi_d \geq p_d$).
In other words, after patients have been followed for $\tau$, we do not care whether they experience DLT eventually.
Nevertheless, the interpretations of $f_{T \mid Z}$ and $S_{T \mid Z}$ are slightly different, as we have $\int_{0}^t f_{T \mid Z}(v \mid Z_i = d, p_d, \xi_d, \bw) \, \dd v  \leq \xi_d $ for any finite $t$.

\end{document}